\documentclass[reprint,eqsecnum,floats,aps,amsmath,amssymb,nofootinbib,prd,onecolumn,showpacs]{revtex4-2}

\usepackage{graphicx}
\usepackage{bm}
\usepackage{amsmath}	
\usepackage{amsfonts}
\usepackage{mathtools}
\usepackage{amssymb}							
\usepackage{braket}
\usepackage{amsthm}			
\usepackage{hyperref}

\usepackage[T1]{fontenc} 

\usepackage{color}

\begin{document}

\title{Comparing Analytic and Numerical Studies of Tensor Perturbations in Loop Quantum Cosmology}

\author{Guillermo A. Mena Marug\'an}
\email{mena@iem.cfmac.csic.es}
\affiliation{Instituto de Estructura de la Materia, IEM-CSIC, Serrano 121, 28006 Madrid, Spain}

\author{Antonio Vicente-Becerril}
\email{antonio.vicente@iem.cfmac.csic.es}
\affiliation{Instituto de Estructura de la Materia, IEM-CSIC, Serrano 121, 28006 Madrid, Spain}

\author{Jes\'us Y\'ebana Carrilero}
\email{jesus.yebana@uib.es}
\affiliation{Departament de Física, Universitat de les Illes Balears, IAC3 – IEEC, Crta. Valldemossa km 7.5, E-07122 Palma, Spain}

\begin{abstract}
We investigate the implications of different quantization approaches in Loop Quantum Cosmology for the primordial power spectrum of tensor modes. Specifically, we consider the hybrid and dressed metric approaches to derive the effective mass that governs the evolution of the tensor modes. Our study comprehensively examines the two resulting effective masses and how to estimate them in order to obtain approximated analytic solutions to the tensor perturbation equations. Since Loop Quantum Cosmology incorporates preinflationary effects in the dynamics of the perturbations, we do not have at our disposal a standard choice of privileged vacuum, like the Bunch--Davies state in quasi-de Sitter inflation. We then select the vacuum state by a recently proposed criterion which removes unwanted oscillations in the power spectrum and guarantees an asymptotic diagonalization of the Hamiltonian in the ultraviolet. This vacuum is usually called the NO-AHD (from the initials of Non-Oscillating with Asymptotic Hamiltonian Diagonalization) vacuum. Consequently, we compute the power spectrum by using our analytic approximations and by introducing a suitable numerical procedure, adopting in both cases an NO-AHD vacuum. With this information, we compare the different spectra obtained from the hybrid and the dressed metric approaches, as well as from the analytic and numerical procedures. In particular, this proves the remarkable accuracy of our~approximations.
\end{abstract}

\maketitle

\section{Introduction}

A challenge of  Modern Physics is the construction of a complete and satisfactory quantization of General Relativity (GR). A candidate that has received considerable attention is Loop Quantum Gravity (LQG) \cite{LQG, Thie}. LQG is a nonperturbative quantization program for GR originally based on a canonical formulation of Einstein gravity in terms of the Ashtekar--Barbero variables. These real variables consist of a gauge connection and a densitized triad, which capture the physical information about the spatial geometry and its extrinsic curvature. The application of LQG techniques to symmetry-reduced models, for considering quantum gravitational phenomena in the Early Universe or in Astrophysics, has led to a discipline called Loop  Quantum Cosmology (LQC) \cite{LQC}. A major achievement of LQC is the resolution of the cosmic singularity problem (at least for certain families of quantum states)~\cite{APS,APS1}. LQC replaces the conventional concept of the {\sl  {Big} Bang}\, with a cosmic bounce, commonly referred to as the {\sl Big Bounce}. This bounce signifies a quantum transition, enabling a passage from a contracting phase of the universe to an expanding~one.

In order to falsify any theory of quantum gravity, ultimately it is necessary to confront its predictions with observational data. In cosmology, the Very Early Universe provides a potential arena where various theories can be tested. In particular, tensor cosmological perturbations may contain information about these stages of the universe. Two observables related to tensor perturbations are the B-mode polarization power spectrum extracted from the Cosmic Microwave Background (CMB)  \cite{Planck} and the Cosmological Background of Gravitational Waves (CBGW), seeded by primordial perturbations originated in the early periods of the universe, and which has yet to be observed \cite{CBGW, LISA} (although a CBGW has been detected by the NANOGrav Collaboration, of which the origin remains unclear~\cite{NANOGrav, NANOGrav2}). In other observables associated with the CMB, such as the distribution of temperatures which gives rise to the angular power spectrum, the polarization of E-modes, or the TE cross-correlation spectrum, the main contribution is due to scalar (rather than tensor) cosmological perturbations. Some of these observables have not yet been measured with sufficient precision, and others, such as the angular power spectrum, exhibit anomalies which may need an explanation from Physics that departs from standard inflationary scenarios~\cite{ASr,ASr2,AgSr,AgSr2}. 

In this paper, we focus our attention on tensor perturbations, which are simple to analyze, with manageable propagation equations. Scalar perturbations have already been addressed in Ref. \cite{Paper_2}, reflecting their different dynamical behavior and their potentiality for a more direct contact with available observational data.

In standard inflationary scenarios, primordial tensor perturbations originate in a vacuum state adapted to the background dynamics and evolve with the expansion \cite{Mukhanov1}. It is common to consider quasi-de Sitter expansion and therefore adopt a slow-roll approximation to inflation. Then, a preferred vacuum state emerges for the perturbations, namely the Bunch--Davies state \cite{Mukhanov1, Bunch}. It is the unique Hadamard state that is invariant under the de Sitter isometry group. In this setting, perturbations can be quantized treating their modes as generalized quantum harmonic oscillators \cite{Baumann, Langlois}. However, this scenario fails to incorporate the effects of preinflationary periods departing from quasi-de Sitter expansion, at least for modes with wavelengths in these periods of the order of the characteristic scale of such effects. In fact, this happens in LQC, with preinflationary dynamics which differ radically from a de Sitter evolution. Therefore, cosmological perturbations cannot be treated quantum mechanically as in slow-roll models. Several approaches have been proposed for the quantization of these primordial perturbations within LQC \cite{hybr_rev,dressed1,dressed2,effective2,effective3,effective4,effective5}. In this work, we will concentrate our attention on two specific approaches that exhibit a good ultraviolet behavior, similar to that found in standard inflationary scenarios. These are the hybrid \cite{hybr_ref,hybr_ten,hybr_rev} and the dressed metric approaches \cite{dressed1,dressed2,dressed3,Agullo1}. 

While there are notable similarities between these two methods, their disparities become crucial during periods with relevant quantum corrections, when the background evolution deviates from standard relativistic  dynamics. This discrepancy arises from the distinct strategies employed in the two approaches to the quantization of the perturbations. The hybrid approach rests on a canonical quantization of the entire system, formed by the homogeneous geometry and its perturbations. In contrast, the dressed metric approach incorporates the most important quantum gravitational effects on the homogeneous background by  means of a dressed metric, then describing the evolution of the perturbations as test fields which propagate in this dressed background. In the regime of effective LQC for the homogeneous geometry, and disregarding any backreaction, these two strategies yield different time-dependent masses for the equations of motion of the perturbations when they are written in a generalized harmonic oscillator form \cite{NBMmass}. 

To fix the initial conditions for the perturbations, we must specify their quantum state, normally interpreted as a vacuum. As we have commented, the identification of this vacuum with the Bunch--Davies state loses significance when we consider preinflationary periods with typical scales that can evolve into those observed today in the CMB. In these situations, we need alternative criteria to determine an appropriate vacuum state. Different criteria yield different solutions, resulting in general discrepant observational predictions \cite{NBM}. To address this issue, various criteria have been proposed in the literature for selecting viable vacuum candidates. One commonly used prescription is the adoption of adiabatic states \cite{Parker}. Other proposals have explored strategies such as minimizing quantities associated with the renormalized stress--energy tensor \cite{Agullo1, Lueders, Handley}, ensuring minimal quantum uncertainty around the bounce while maintaining classical properties at the end of inflation~\cite{AG1, AG2}, or minimizing power oscillations, either during a period that extends from a kinetically dominated epoch until the beginning of inflation \cite{deBlas}, or by means of an asymptotic diagonalization of the Hamiltonian of the perturbations in the ultraviolet sector \cite{NMT}. In this work, we are going to adhere to this last proposal to determine the vacuum state of the perturbations. Indeed, it has been demonstrated that when requiring an asymptotic Hamiltonian diagonalization, one can obtain spectra with non-oscillating behavior, eliminating spurious contributions to the predicted power. Moreover, the asymptotic diagonalization allows for a natural choice of positive frequencies in the ultraviolet, ensuring the good properties that one typically expects of adiabatic states when these are well defined \cite{NMT}. In particular, the vacuum selected with this criterion reproduces the standard Poincar\'e vacuum in Minkowski spacetime, and the Bunch--Davies state for de Sitter cosmologies~\cite{NMT}. Herein, we focus our investigation on these types of vacuum states that are well adapted to quantum effects, and we will refer to a non-oscillating (NO) vacuum determined by asymptotic Hamiltonian diagonalization (AHD) as an NO-AHD vacuum.

Our aim in this work is to compute the primordial power spectrum (PPS) of tensor cosmological perturbations using two different quantization approaches within the framework of LQC, namely the hybrid and the dressed metric approaches. We will conduct this computation in two different ways. Firstly, by analytic approximations, which will allow us to derive and handle explicit expressions containing the parameters of our model, and secondly by numerical integration, something that has never been done before for tensor modes in this vacuum state. The goal is to complete both computations in order to compare them and thus verify whether our analytic approximations adjust well with the exact numerical solution. We will show that the agreement is remarkably good. Our analytic approximations are based on some recently published developments in LQC \cite{NM}, which we here extend and improve by various means, such as the inclusion of slow-roll corrections in the purely inflationary era. Our intention is to present a comprehensive discussion of the treatment of tensor perturbations in LQC, including a succinct explanation of the main features of the two discussed quantization approaches. Our analysis is a cornerstone for the parametrization of the PPS in terms of the background initial conditions and the constants of the model that affect its dynamics, taking as a starting point our approximations once they are validated by numerical studies. 

The structure of this paper is as follows. In Section \ref{sec2}, we briefly review and numerically compute the dynamical equations of the background, described by effective LQC~\cite{LQC,taveras}. In Section \ref{sec3}, we introduce the effective masses of the tensor perturbations for the hybrid and dressed metric approaches \cite{NBMmass}. In doing so, we summarize the main points of the derivation of the propagation equations for the tensor modes in the two considered quantization approaches. We then define NO-AHD vacua in Section \ref{sec4} and employ this definition to determine the initial state of the tensor perturbations. Since the mode equations cannot be solved exactly, in Section \ref{sec5} we approximate the effective mass in a convenient form that allows us to reach the desired analytic resolution. Then, in Section \ref{sec6} we calculate analytically the PPS. We also present the numerical integration procedure that we follow to compute the exact power spectrum. Additionally, in Section \ref{sec6} we compare the different power spectra, discussing the differences between them and their robust features. This comparison is made both between the spectra corresponding to the two approaches within LQC and between the analytic and numerical results. Finally, Section \ref{sec7} contains the conclusions. In the following, we use Planck units, setting $G$, $c$, and $\hbar$ equal to one.

\section{Background Dynamics}\label{sec2}

We consider a homogeneous and isotropic Friedmann--Lemaître--Robertson--Walker (FLRW) spacetime with a homogenous scalar field $\phi(t)$, which plays the role of the inflaton. This field is subjected to a potential $V(\phi)$. For concreteness, we will focus our attention on a quadratic potential $V(\phi)=m^2\phi^2/2$, where $m$ is a constant mass. However, it is easy to see that our analysis can be generalized to other potentials without major obstructions. The quadratic potential has been studied in detail in LQC and allows for an easier implementation and a direct interpretation of the results. Even though this potential has been disfavoured by observations in a standard cosmological model within GR \cite{PlanckInfla}, these considerations are not straightforwardly applicable to LQC, where a best fit of cosmological parameters including the inflaton potential has not been carried out yet in full detail. In addition, for simplicity, we assume a flat spatial topology with compact spatial sections isomorphic to the three-torus $\mathbb{T}^3$, possessing a compactification length of $l_0$ (the non-compact case can be attained in a convenient limit in which $l_0$ is removed). The FLRW spacetime metric depends on the scale factor $a(t)$ and the homogeneous lapse $N_0(t)$. This metric is given by 
\begin{equation}
ds^2 = - N_0^2(t) dt^2 + a^2(t)\; ^0h_{ij}dx^idx^j ,
\end{equation}
where $i,j=1,2,3$ denote spatial indices, $^0h_{ij}$ is the Euclidean three-metric on the compact spatial section, and $x^i$ are periodic Euclidean coordinates, with a period equal to $2\pi/l_0$. The quantization of this homogeneous and isotropic model, without potential, has been thoroughly studied in LQC \cite{APS1,APS_Ham}. In particular, it is possible to construct a well-defined operator for its Hamiltonian constraint \cite{APS1,hybr_rev}. Although this construction is not free of regularization ambiguities \cite{amb1,amb2,amb3}, it has been proven that this quantum Hamiltonian is essentially unique under certain requirements of possessing a minimal number of terms~\cite{Engle}. Among the solutions to this quantum constraint, there exist states that are highly peaked on trajectories that are generated by a specific effective Hamiltonian, differing from the classical one by the incorporation of quantum corrections. These trajectories avoid the classical cosmological singularity and instead bounce \cite{LQC,APS1,taveras}, connecting a contracting branch to an expanding branch of the universe. In the presence of a nonvanishing potential, it has been shown numerically that the effective behavior is not qualitatively modified, also displaying peaked states with a quantum bounce  \cite{ads,lambd}. Moreover, even other Hamiltonian regularizations have been seen to lead to similar background dynamics in the expanding cosmological branch after the bounce \cite{Wang1,Wang2}. The effective Hamiltonian is 
\begin{eqnarray}
N_0 H_{|0}^{eff} = \frac{N_0}{2l_0^3 a^3} \left[\pi_\phi^2 - \frac{3l_0^6 a^6}{4\pi \gamma^2 \Delta} \sin^2{\left(\frac{4\pi \gamma \sqrt{\Delta}\pi_a}{3l_0^2a^2} + 2 a^6 l_0^6 V(\phi)\right)} \right],
\end{eqnarray}
where $\pi_\phi$ and $\pi_a$ are, respectively, the canonically conjugate momenta of $a$ and $\phi$. In addition, $\gamma$ is the so-called Immirzi parameter \cite{immirzi} and $\Delta = 4 \sqrt{3}\pi \gamma $ is the area gap allowed by the spectrum of the area operator in LQG \cite{LQG,Thie}. As is usual in LQG, we set $\gamma=0.2375$, a value supported by black hole entropy calculations \cite{BHLQG1,BHLQG2}. The above effective LQC Hamiltonian yields the dynamical equations that govern the background (see, e.g., Refs. \cite{dressed3,NBM}),
\begin{eqnarray}\label{eq_Friedman_LQC}
    \left( \frac{a'}{a}\right)^{2} = \frac{8\pi}{3}a^{2}\rho \left(1 - \frac{\rho}{\rho_{c}} \right),\quad \frac{a''}{a} = \frac{4\pi}{3}a^{2}\rho \left(1 + 2\frac{\rho}{\rho_{c}} \right) - 4\pi a^{2}P \left(1 - 2\frac{\rho}{\rho_{c}} \right).
\end{eqnarray}
Here, the prime denotes the derivative with respect to the conformal time while, in the following, we will use a dot to denote the proper-time derivative. The energy density $\rho$ and the pressure $P$ of the scalar field are 
\begin{eqnarray}
    \rho = \frac{1}{2} \left(\frac{\phi'}{a} \right)^2
    + V(\phi)  , \quad P =\rho - 2 V(\phi).
\end{eqnarray}
In addition, $\rho_c = 3 /(8\pi \gamma^2 \Delta)$ is the maximum that the energy density $\rho$ can reach. It is attained at the bounce, and it is usually called the critical density. 

To simplify the comparison between the analytic and numerical studies that we want to carry out, without the need for large numerical resources and computation times, in the following we fix the mass parameter of the quadratic inflaton potential to a convenient value, $m=1.2 \times 10^{-6}$. This value lies well within an interval of parameters $m$ that are phenomenologically favored in order to have a sufficient number of e-folds and attain predictions that are compatible with the observations of the CMB, while still allowing for non-negligible quantum modifications with respect to standard slow-roll inflation \cite{dressed3}. Such predictions do not vary much if $m$ is slightly changed. On the other hand, it is worth noticing that, in our analytic investigations, we might leave $m$ as a free parameter of the model, running in the commented interval, and derive formulas for the primordial spectra that would explicitly depend on it. We assume that, if our analytic fitting to the numerical power spectrum is satisfactory for the chosen value of $m$, this will continue to be so for values around it.

\begin{figure}
    \includegraphics[width=16cm]{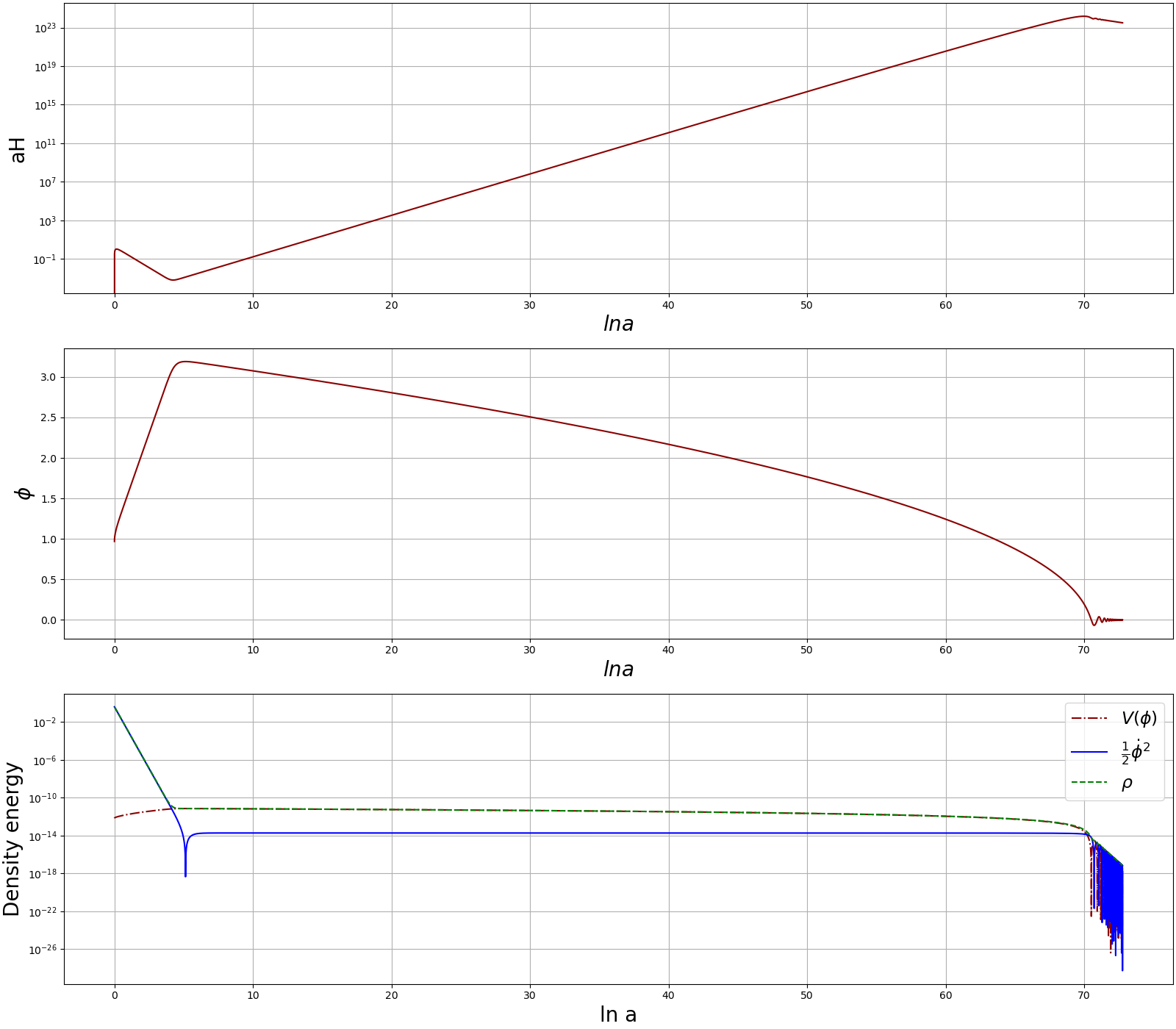}

    \caption{From top to bottom, evolution of the inverse of the co-moving Hubble radius, the scalar field, and the kinetic, potential, and total energy densities. This evolution is displayed in terms of the number of e-folds (rather than conformal time).}
    \label{fig_Background}
\end{figure}

Let us numerically study the dynamical equations of the background. For this purpose, we will use the Friedmann Equation \eqref{eq_Friedman_LQC} and the local conservation law
\begin{eqnarray}
    \Ddot{\phi} + 3 H \dot{\phi} + V_{,\phi} = 0,
\end{eqnarray}
where $H=a^{\prime}/a^2$ is the Hubble parameter.~We have integrated this system of (two) differential equations using a fourth-order Runge--Kutta algorithm. We have adopted initial conditions which have frequently been chosen in the LQC literature (for phenomenological reasons similar to those explained above when we fixed $m$) \cite{dressed3,NBM,NM}. Selecting these initial conditions makes it possible to compare our study with previous results in LQC. We impose these conditions at the bounce time $t=t_B$, where the Hubble parameter identically vanishes according to effective LQC, namely $H (t_B)=0$. Moreover, we take $a(t_B) = 1$ as a reference scale. Then, the only initial value that we have to fix is that of the inflaton, for which we choose $\phi(t_B) = 0.97$. With these data, and using the Hamiltonian constraint, it is easy to compute the time derivative of the inflaton at the bounce. After numerical integration, we obtain the background evolution displayed in Figure \ref{fig_Background}.

We have split this numerical integration into two different time intervals. First, from $t_B$ (considered as the origin of time) to an intermediate instant at $10^2$, and second from this instant to a time when inflation has already ended, $t_{fin}=10^8$. For each of these intervals, we have used $10^8$ points with a uniform point distribution. Our procedure is motivated by the technical requirements necessary for a satisfactory precision both during the period where the quantum effects are non-negligible and during the fast evolution experienced in the inflationary period. Our numerical results coincide, e.g., with those obtained in Ref.~\cite{NBM}. 

We can see in Figure \ref{fig_Background} that the co-moving Hubble radius decreases rapidly during inflation (i.e., its inverse increases). In addition, note that the expansion reaches around $70$ e-folds, indicating a short-lived inflation that still satisfies the observational bounds necessary to solve the horizon and flatness problems \cite{Liddle}. This total number of e-folds from the bounce until the end of inflation is in agreement with other numerical computations performed with similar values of the mass parameter and the initial condition on the inflaton, for instance in Ref. \cite{AG1}. As the co-moving radius decreases, all visible modes would cross the horizon, when $k=aH$ \cite{Langlois, Baumann}, and become frozen \cite{NBM}. Nonetheless, we see that modes with wavenumbers approximately in the interval $10^{-1}\le k \le 1$ enter and exit the horizon even before inflation. These modes are the first that freeze and their propagation inside the horizon takes shorter times. We will pay special attention to these modes in our later study of the PPS, although we will analyze a much wider window of modes to include a much richer variety of physical phenomena.

On the other hand, the numerical computation suggests that all observable modes cross the horizon and freeze after approximately $30$ e-folds \cite{NM}.~This implies that the PPS can be evaluated at these stages of the evolution, when the relevant modes have frozen, instead of unnecessarily extending the numerical calculations to integrate the mode dynamics up to later times. 

\section{Effective Mass in the Hybrid and Dressed Metric Approaches}\label{sec3}

Let us briefly overview the quantization procedures followed in the hybrid and dressed metric approaches. While this section is not essential for the understanding of the article, it contributes to a self-contained discussion. Readers unfamiliar with LQC techniques can opt to skip it. We do not attempt to repeat here the specific details of the two considered quantization approaches, which can be found, e.g., in Refs. \cite{NBMmass,hybr_rev}, but rather provide a succinct explanation of the distinct strategies followed in each case and clarify the origin of the differences between the two expressions of the effective mass. 

\subsection{Hybrid Quantization Approach}

The hybrid approach \cite{hybr_ref,hybr_rev} entails a simultaneous quantization of the background and the perturbations as a constrained canonical system, starting from the gravitational action in Hamiltonian form and truncated at the second perturbative order. Focusing exclusively on tensor perturbations, the corresponding total Hamiltonian is given by the sum of the background Hamiltonian, which would be constrained to vanish in the absence of perturbations, and a quadratic contribution of the tensor perturbations \cite{hybr_rev}. In this tensor case, there are no perturbative diffeomorphism constraints, and the perturbations are directly gauge invariants. We choose canonical variables for the tensor perturbations that render the associated (Fourier) mode equation into a generalized harmonic oscillator equation, without a friction term \cite{hybr_rev}. We call $d_{\Vec{k}}^{(\epsilon)}$ the configuration coefficients of the tensor gauge invariants in the mode expansion, where $\Vec{k}$ is the wavevector and $\epsilon$ represents the polarization. The canonical momenta of these variables are also gauge-invariant. The ambiguity in adding to them a linear term in the configuration variables can be fixed by requiring a Hamiltonian without mixed configuration--momenta terms, or equivalently by demanding that the time derivative of the momenta be proportional to the corresponding configuration variables. 

The obtained canonical system is only subjected to the constraint that the total Hamiltonian must vanish. One can then quantize it by adopting a loop representation for the (square) scale factor, a standard quantum mechanical representation for the inflaton, and a Fock representation for the tensor perturbations (see Ref. \cite{hybr_rev} for further details). The quantization of the contribution of the perturbations to the total Hamiltonian is achieved by adopting the same operator representation for the square scale factor and its momentum as in the background term, while the quadratic factors in the perturbations are represented in terms of annihilation and creation operators \cite{hybr_rev}, using normal ordering. Products of (positive) functions of the square scale factor and its momentum are symmetrized algebraically. The most important feature of this hybrid quantization as far as the effects on the perturbations are concerned is the fact that the extrinsic curvature of the geometry is encoded in terms of a canonical (momentum) variable, instead of defining it dynamically. If one then considers quantum states with a separated dependence on the (square) scale factor, on the one hand, and the perturbations, on the other hand, and assuming that the perturbations do not mediate changes in the FLRW geometry, one can arrive at a master constraint for the tensor perturbations by a kind of mean-field approximation. This approximation identifies as the most relevant part of the constraint for the perturbations just the expectation value over the geometric dependence, evaluated on the part of the quantum state that varies with the scale factor \cite{hybr_rev}. Note that this expectation value must be computed using the inner product of LQC. For quantum states that are peaked, the expectation values should coincide with the evaluation of the canonical geometric variables on the peak trajectories. In particular, there exist states for which the backreaction is negligible, even compared to the quantum corrections of the background around the bounce, and their peaks follow the dynamics of effective LQC, on which we focus the rest of our discussion. For them, the expectation values of the geometry should reproduce the evaluation of the (square) scale factor and its momentum in effective LQC. Then, the propagation equations of the tensor modes adopt the expression  
\begin{equation}\label{eq_MS_tens_hyb}
    d_{\Vec{k}}^{(\epsilon)''} + \left[k^2 - \frac{4\pi }{3} a^2 (\rho-3P)\right] d_{\Vec{k}}^{(\epsilon)} = 0.
\end{equation}
Here, $k$ is the wavenumber, equal to the (Euclidean) norm of the wavevector $\Vec{k}$. We call the term in square brackets the effective mass with the mode-dependent contribution $k^2$ removed. It is worth emphasizing that this effective mass differs from its counterpart in GR, namely $-a''/a$. Nonetheless, a rapid inspection of Equation \eqref{eq_MS_tens_hyb} and the second identity in Equation \eqref{eq_Friedman_LQC} shows that one indeed recovers the GR dynamics for the perturbations in the limit $\rho_c \rightarrow \infty$ or, equivalently, when the energy density becomes much smaller than the critical one.

\subsection{Dressed Metric Approach}

The dressed metric approach \cite{dressed1,dressed2,dressed3,Agullo1} also combines a quantization of the background utilizing LQC techniques and of the perturbations employing a Fock representation. However, in contrast to the hybrid approach, the system is not treated canonically as a whole. The background geometry is quantized first following the standard rules of LQC, completely neglecting backreaction. Once the background metric is dressed with quantum corrections, it is lifted to the phase space of the perturbations, where the dynamics is governed by a quadratic Hamiltonian which is not constrained to vanish, since backreaction is fully disregarded at this level \cite{dressed2,dressed3}. As a consequence of this procedure, the evolution of the dressed metric for peaked states of the background geometry is governed by the effective dynamics of homogeneous and isotropic LQC. Therefore, the time derivatives of this dressed metric do not satisfy the Hamiltonian equations of GR that relate them with the canonical momentum of the square scale factor. This immediately implies that the expression of the extrinsic curvature in terms of the time derivatives of the dressed metric differs from its counterpart in terms of the canonical momentum of the geometry, evaluated on effective trajectories. Owing to this discrepancy, we can already anticipate that the effective mass of the tensor perturbations is not the same in the dressed metric and the hybrid approaches.

In more detail, in the dressed metric approach the reduced phase space of the perturbations is usually described with a specific choice of tensor variables (which are gauge-invariant, as explained in the hybrid case). The perturbations are seen as test fields propagating on a dressed metric, which can be described by a number of expectation values that encapsulate the quantum geometry effects on the background. Again, for highly peaked states, these expectation values can be obtained by evaluating the geometry on effective trajectories for LQC. Nonetheless, opposite to the situation described for the hybrid approach, the geometric quantities that are evaluated are, in principle, the scale factor and its conformal time derivatives, rather than functions of the square scale factor and its canonical momentum. Since the relation between this momentum and the derivative of the scale factor is not the same before quantization than in effective LQC, as we have pointed out, the outcome of the evaluation on effective trajectories differs in the two approaches \cite{NBM}. 

The mode equation governing the evolution of the tensor perturbations in the dressed metric approach is \cite{dressed2,dressed3}
\begin{equation}
    \mathcal{T}_{\Vec{k}}^{(\epsilon) ''} + 2 \frac{a'}{a} \mathcal{T}_{\Vec{k}}^{(\epsilon)'} + k^2 \mathcal{T}_{\Vec{k}}^{(\epsilon)} = 0. \label{eq_DS_Tensor_original} 
\end{equation}
We adopt the notation of Ref. \cite{dressed2}, with $\mathcal{T}_{\Vec{k}}^{(\epsilon)}$ denoting the configuration coefficient of the tensor mode (apart from a factor of $l_0^{-3}$), where $\epsilon$ is again the polarization label. The above expression coincides formally with the classical result obtained in GR for an expanding universe, except that the time derivative of the scale factor must be evaluated using effective LQC. To enable a comparison between the dressed metric and the hybrid mode equations, we can perform the change of variables 
\begin{equation}
    {\bar{d}}_{ \Vec{k}}^{(\epsilon)} = \frac{a}{\sqrt{32\pi l_0^3}} \mathcal{T}_{ \Vec{k}}^{(\epsilon)},
\end{equation}
where $a$ corresponds to the effectively dressed scale factor. This change leads to the counterpart of Equation \eqref{eq_MS_tens_hyb}: 
\begin{equation} \label{eq_MS_tens_dress}
    {\bar{d}}_{ \Vec{k}}^{(\epsilon)''} + \left[k^2 - \frac{4\pi}{3}a^{2}\rho \left(1 + 2\frac{\rho}{\rho_{c}} \right) + 4\pi a^{2}P \left(1 - 2\frac{\rho}{\rho_{c}} \right) \right]{\bar{d}}_{ \Vec{k}}^{(\epsilon)} = 0 .
\end{equation}
with ${\bar{d}}_{ \Vec{k}}^{(\epsilon)}$ playing the same role in the dressed metric approach as $d_{ \Vec{k}}^{(\epsilon)}$ in the hybrid formalism. In the regime in which the background effective dynamics coincide in practice with the classical dynamics of GR, this equation reproduces the relativistic equation for the propagation of tensor perturbations. This can be seen by noticing that, in the limit where $\rho_c$ becomes large (or the energy density becomes much smaller than it), we recover Equation \eqref{eq_MS_tens_hyb}, which in this limit converges to the GR dynamics.

\begin{figure}

    \includegraphics[width=16cm]{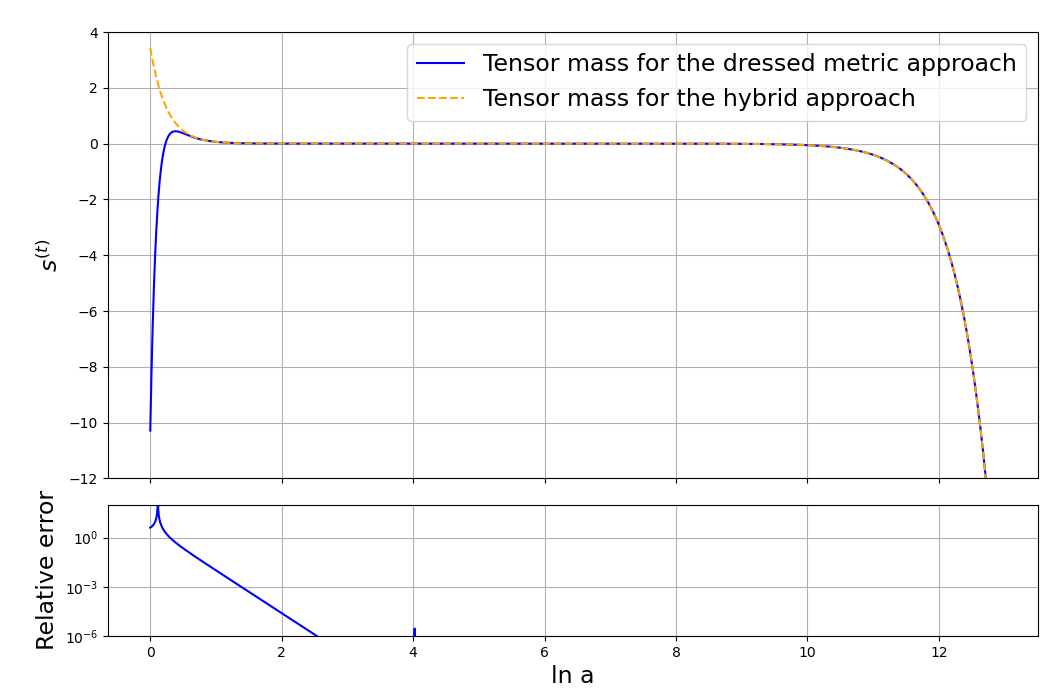}

    \caption{\textbf{Top}: Numerical computation of the tensor background-dependent mass $s^{(t)}$ in the dressed metric approach (blue solid line) and the hybrid approach (dashed orange line). These background-dependent masses are defined in Equation \eqref{eq_MS_tens_hyb} and Equation \eqref{eq_MS_tens_dress}, respectively. \textbf{Bottom}: Relative error between the value of the tensor mass in the dressed metric approach and the hybrid approach. We see that both masses become indistinguishable in the inflationary period. We have taken $\gamma=0.2375$ and $\phi_0=0.97$ for the Immirzi parameter and the value of the inflaton at the bounce, respectively, and considered a quadratic inflaton potential with mass $m=1.2 \times 10^{-6}$.}
    \label{fig_difference}
\end{figure}

The difference between the effective mass in the hybrid and dressed metric approaches is evident at the bounce, where the mass is positive in the hybrid case (at least for phenomenologically interesting backgrounds with small contribution of the inflaton potential), while it is always negative for the dressed metric case, because the effective trajectories have a minimum in the scale factor (so that $-a''/a<0$). For our initial data and potential, we also see in Figure \ref{fig_difference} that the effective mass in the dressed metric approach changes its sign and then remains positive until the onset of inflation, whereas the mass of the hybrid approach is strictly positive during the whole evolution until inflation. Note also that, in the inflationary period, where the quantum geometry effects are negligible, the two effective masses become indistinguishable.

\section{Initial Conditions and Vacuum State} \label{sec4}

In order to solve the propagation equation of the tensor modes and compute them at the end of inflation, either analytically or numerically, it is essential to know their initial conditions. These conditions fix the initial state of the perturbations, and vice versa. It is most reasonable to identify this state with a  vacuum for the perturbations. Nonetheless, the question arises of what a vacuum state is in curved, nonstationary spacetimes, like those considered in this work. In the slow-roll regime of standard inflationary cosmology, the Bunch--Davies state is the most natural vacuum, since it is the unique Hadamard state that is invariant under the de Sitter isometry group, leading to a non-oscillatory and quasi-invariant PPS \cite{Mukhanov1,Bunch}. But, in our case, where preinflationary epochs are involved, the Bunch--Davies state is not a privileged choice, because it is not well adapted to the background dynamics, at least for perturbation modes able to feel the quantum geometry effects around the bounce. In these scenarios, it is necessary to specify and select a different preferred quantum state as a vacuum for the gauge-invariant modes. With this aim, several proposals have been put forward in the context of LQC \cite{Agullo1,Lueders,Handley,AG1,AG2,deBlas,NMT}.

A frequent choice of vacuum state is given by the so-called adiabatic states \cite{Parker,Lueders}. These states were originally introduced as approximated solutions in cosmological spacetimes with slow variation regimes and have a particularly good ultraviolet behavior similar to that of plane waves in Minkowski spacetime. They also possess good regularization properties, at least for adiabatic states of a sufficiently high order \cite{Parker1,Anderson}. However, their definition depends on the order of adiabatic iteration in their construction and, perhaps more importantly, also on the initial time when they are determined. The authors of Ref.~\cite{Agullo1}, for instance, proposed to select the unique state that provides a vanishing regularized stress--energy tensor (mode by mode) at the given initial time, but the dependence on this time remains. Moreover, adiabatic choices in the cosmological branch previous to the bounce are affected by regularization ambiguities (because the background dynamics changes considerably for this branch, see Ref. \cite{Wang1}), whereas the adiabatic approximation breaks down around the bounce for small wavenumbers if the effective mass is negative there, as happens in the dressed metric approach \cite{NBMmass}.

Ashtekar and Gupt \cite{AG1,AG2} proposed to fix the vacuum by requiring a subtle interplay between the behavior in the region with important LQC effects (with an energy density of only a few orders below Planck scale) and the behavior at the end of inflation. In the quantum region, the Weyl curvature of the state must remain below a certain bound, compatible with the uncertainty principle. At the end of inflation, the state should display a classical behavior with certain properties (in the case of the scalar perturbations discussed in Ref. \cite{AG1}, it must minimize the square of the Ricci tensor of the spatial metric). Another proposal for a vacuum state is the so-called non-oscillating vacuum introduced in Ref. \cite{deBlas}. This vacuum is characterized by minimizing the oscillations in the square norm of the perturbations when integrated over certain evolution interval. In this work, the suggested interval covered the period from the bounce to the beginning of inflation. An independent proposal which may be related to this previous one consists of choosing a state of low energy that minimizes the regularized energy density distributed along the time-like curve of an isotropic observer \cite{SLE}. Finally, another proposal that selects a vacuum state without large oscillations in (the norm of) their amplitude is the NO-AHD proposal \cite{NMT,NM}, which we will describe in full detail later in this section.

Based on the study of all these proposals, during recent years it has become evident that the choice of vacuum state has a great impact on the PPS. In particular, as we have partly commented, the spectrum displays an oscillatory behavior for generic choices of this vacuum, at least for modes with a small wavenumber. This is because the chosen state is not optimally adapted to the evolution of the modes, with rapid variations of phase. These oscillations, when binned, usually result in an average power enhancement, which can be considered artificial since we can eliminate it with a judicious change of state. The corresponding non-oscillatory power spectrum will still differ from the spectrum obtained in standard relativistic inflation by quantum effects introduced by LQC during the preinflationary epoch (see, e.g., the discussion in Ref. \cite{NM}). 

On the other hand, we note that most of the proposals for the choice of a vacuum for the perturbations within LQC depend strongly on the time at which initial conditions are set and/or need numerical computations, either to minimize a quantity associated with the stress--energy tensor \cite{Agullo1,Lueders,Handley,SLE}, to ensure classical properties at the end of inflation among a family of states with minimal quantum uncertainty around the bounce \cite{AG1,AG2}, or to minimize power oscillations during a period that covers a kinetically dominated epoch extending until the beginning of inflation \cite{deBlas}. A choice of vacuum state that avoids these extra complications, and in principle can be determined without resorting to numerics, is given by the NO-AHD proposal \cite{NMT}. This is the proposal that we will follow in this work. NO-AHD states are picked out by the criterion of the (ultraviolet) asymptotic diagonalization of the Hamiltonian of the perturbations, guaranteeing a satisfactory adaptation to the background dynamics, as implied by the fact that the mode evolution reduces asymptotically to a rotating phase. Furthermore, beyond the asymptotic region, this vacuum state ultimately provides us with a non-oscillatory behavior in the PPS, avoiding any spurious increase in power. Actually, it has been observed that this vacuum leads to power suppression at a physical scale corresponding at the bounce to the curvature scale there \cite{NM}. This suppression could be a favorable option to minimize the tensions observed for low multipoles in the angular spectrum of the CMB with respect to the standard inflationary predictions, in the case of scalar perturbations.

As we have commented, the NO-AHD proposal selects a vacuum state that can be determined analytically, at least in cosmological situations like those studied in this work~\cite{NM}. Indeed, we can fix this vacuum state at the bounce where the effective mass adjusts very well with a Pöschl--Teller (PT) potential, as will be seen in the next section. Evolving it, we can then specify the desired state at any other time. In more detail, any (normalized) solution to the mode equation can be expressed in the form
\begin{eqnarray} \label{NO-AHD modes}
    \mu_{k} = \sqrt{-\frac{1}{2 \textrm{Im}(h_{k})}}e^{i \int_{\eta_{0}}^{\eta}d\tilde{\eta} \textrm{Im}(h_{k})(\tilde{\eta})},
\end{eqnarray}
where $\eta$ is the conformal time and $h_{k}$ is a time-dependent function satisfying the Riccati equation
\begin{eqnarray} \label{Riccati}
    \dot{h}_{k} = k^{2} + s^{(t)} + h_{k}^{2}.
\end{eqnarray}
Here, $s^{(t)}$ denotes the effective mass of the tensor perturbations. The sought NO-AHD vacuum is specified by a function $h_{k}$ with the following asymptotic expansion for a large $k$:
\begin{equation}\label{asymptotich} 
    \frac{1}{h_k}\sim  \frac{i}{k}\left[1-\frac{1}{2k^2}\sum_{n=0}^{\infty}\left(\frac{-i}{2k}\right)^{n}\gamma_n \right].
\end{equation}
The coefficients $\gamma_{n}$, independent of the wavenumber, are determined by the recurrence relation
\begin{eqnarray}
    \gamma_{0} = s, \hspace{1.5cm} \gamma_{n+1} = -\dot{\gamma}_{n} + 4s\left[ \gamma_{n-1} 
     \sum_{m=0}^{n-3} \gamma_{m}\gamma_{n-(m+3)}\right] - \sum_{m=0}^{n-1} \gamma_{m}\gamma_{n-(m+1)}.
\end{eqnarray}
It is possible to show that the resulting $h_{k}$ must have a negative imaginary part (at least asymptotically), so that Equation \eqref{NO-AHD modes} provides positive-frequency modes \cite{NMT}. Moreover, the asymptotic expansion determines a unique solution for all $k$ in regimes where the inflaton potential is negligible or sufficiently small \cite{NM,NMY}. This solution will provide the desired initial conditions for our analytic integration.

\section{Approximate Effective Mass and Mode Solutions for the Hybrid and Dressed Metric~Approaches}\label{sec5}

In this section, we will compute analytically the mode solutions of the tensor perturbations at the end of inflation, both for the hybrid and the dressed metric approaches. Rather than exact, we will obtain approximate solutions, derived by describing the effective mass with some approximations \cite{NM, NMY, AMV}. 

\subsection{Hybrid Quantization}

For convenience (for readers who skipped Sec. III), we recall that the effective mass of the tensor perturbations $s^{(t)}$ is given in the hybrid approach by Equation \eqref{eq_MS_tens_hyb},
 \begin{eqnarray} \label{eq_hyb_mass}
     s^{(t)} = - \frac{4\pi G}{3}a^{2}(\rho - 3P).
 \end{eqnarray}
The mode equation with this effective mass cannot be solved analytically. To obtain analytic expressions, we will introduce a series of approximations. We distinguish three different regimes during the evolution of the perturbations. In each of them, we will approximate the effective mass in such a way that the we can proceed analytically.
 
During the bounce epoch, we can describe the effective mass with a PT potential. The quantum effects become negligible as soon as we leave the bounce, but we still remain in a kinetically  dominated regime, so that the effective mass coincides in practice with that found in GR with kinetic domination. As time increases, the potential contribution grows and the kinetic term diminishes, until the former dominates and a standard inflationary epoch begins. In this last epoch (or at least for most of it), the effective mass is well described by an inflationary slow-roll approximation.

\subsubsection{Pöschl-Teller Approximation}

\begin{figure}
    
    \includegraphics[width=16cm]{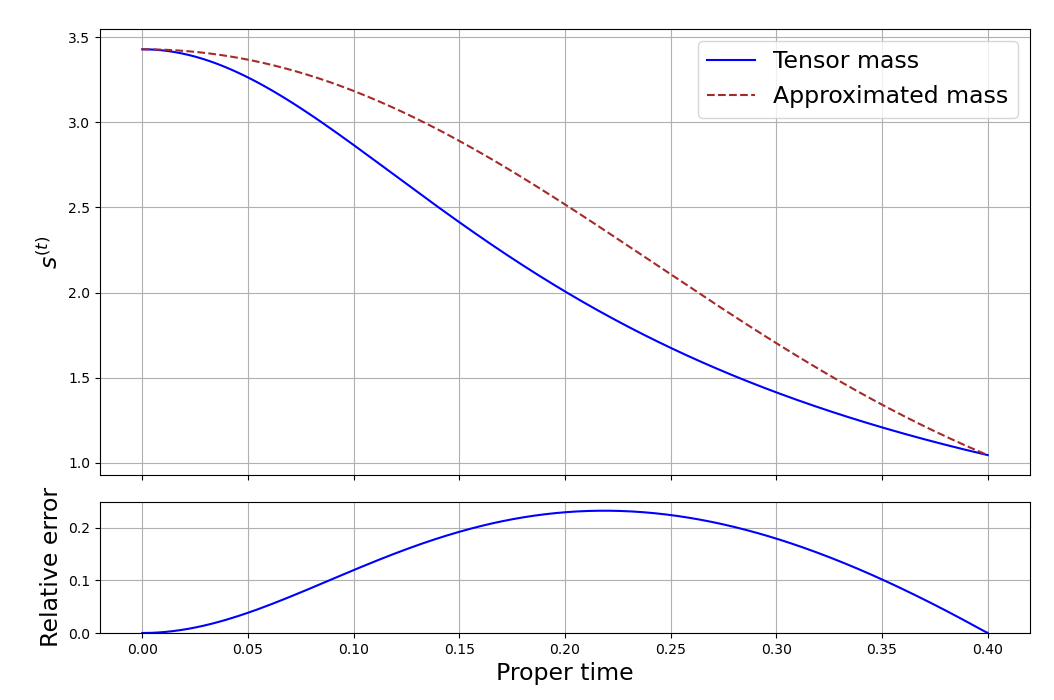}

    \caption{\textbf{Top}: Numerical computation of the tensor mass \eqref{eq_hyb_mass} in kinetic domination in the hybrid approach and the approximated mass \eqref{pth} using the Pöschl--Teller approximation, taking $t_0=0.4$. \textbf{Bottom}: Relative error between the numerical and the approximated values of the mass for tensor perturbations in the hybrid approach. We have taken $\gamma=0.2375$ and $\phi_0=0.97$ for the Immirzi parameter and the value of the inflaton at the bounce, respectively, and considered a quadratic inflaton potential with mass $m=1.2 \times 10^{-6}$.}
    \label{fig_PT_Kinetic_Hybrid}
    \end{figure}

This first epoch takes place around the bounce. During this part of the cosmological evolution, the quantum effects are important. Moreover, in the effective LQC backgrounds of phenomenological interest, the inflaton potential is totally negligible in this period compared with the kinetic contribution. This fact allows us to solve analytically the background effective dynamics, governed by Equation \eqref{eq_Friedman_LQC}, obtaining
\begin{eqnarray}
    a^6 = a_\text{B}^6 \Bigg[ 1+24\pi\rho_c \Big (t-t_\text{B} \Big)^2 \Bigg ]\label{eq_a_LQC},
\end{eqnarray}  
where we recall that $t_B$ is the bounce time (which in practice can be set to zero by choosing it as the time origin). Using this expression, one can compute the effective mass of the perturbations in proper time, as explained in Ref. \cite{NM}. However, the relation between this time and the conformal one used in the mode equation, which is given by the formula $\eta=\eta_B+\int_{t_b}^t dt/a(t)$, cannot be inverted in an analytic form. Hence, an exact analytic solution of the mode equation is not possible. Nonetheless, the effective mass can be well approximated by a PT potential,
\begin{equation} \label{pth}
         s_{PT}^{(t)} = \frac{U_{0}}{\cosh^{2}\left[\alpha(\eta-\eta_{B})\right]}.  
\end{equation}
This approximation was first suggested in Refs. \cite{waco,waco2} and later improved in Ref. \cite{NM} by modifying the procedure used to fix the potential parameters. According to this new procedure, the constants $U_{0}$ and $\alpha$ are such that the PT mass has the same value as the exact one at the bounce and at the end of the considered interval, $\eta_0$. Calling $a_0$ the value of the scale factor at this end, we obtain 
\begin{eqnarray}
         U_{0} = \frac{8\pi\rho_{c}}{3}, \hspace{1.5cm} \alpha = \frac{\mathrm{arcosh}{(a_{0}^{2})}}{(\eta_{0}-\eta_{B})^{2}}.
\end{eqnarray}
There is a certain freedom in fixing $\eta_0$ and hence the duration of this bounce period. In Ref.~\cite{NM}, it was proposed to fix this freedom by optimizing the relative error in the effective mass (both in this interval and in the subsequent kinetically dominated regime). The best compromise is reached when $t_{0} = 0.4$ in proper time. In the following, we take this value as the end of the period. We have computed numerically the relative error introduced with our approximation. As shown in Figure \ref{fig_PT_Kinetic_Hybrid}, it is always below $25\%$. 

With this PT effective mass, the general mode solution is \cite{NM}
\begin{eqnarray} \label{ukh_tens}
u_{k}^{PT} &=& M_{k}\left[ x \left( 1-x\right) \right]^{\frac{-ik}{2\alpha}} {}_{2}F_{1}\left( b_{1}^{k}, b_{2}^{k}, b_{3}^{k}; x \right) \nonumber
        \\
        &+& N_{k}x^{\frac{ik}{2\alpha}} \left( 1-x \right)^{\frac{-ik}{2\alpha}} {}_{2}F_{1}\left( b_{1}^{k} - b_{3}^{k} + 1, b_{2}^{k} - b_{3}^{k} + 1, 2 - b_{3}^{k}; x \right),
\end{eqnarray}
where $x = \left[ 1 + e^{-2\alpha \left(\eta-\eta_{B}\right)} \right]^{-1}$, and $M_{k}$ and $N_{k}$ are integration constants, which can be determined with a choice of vacuum state. The time-independent parameters appearing in the hypergeometric functions ${}_{2}F_{1}$ are
\begin{eqnarray} \label{b1b2b3}
        b_{1}^{k} = c_{+} - \frac{ik}{\alpha},  \hspace{1cm}  b_{2}^{k} = c_{-} - \frac{ik}{\alpha}, \hspace{1cm} b_{3}^{k} = 1 - \frac{ik}{\alpha},\hspace{1cm} c_{\pm} = \frac{1}{2} \left( 1 \pm \sqrt{1 + \frac{32\pi\rho_{c}}{3\alpha^{2}}} \right).
\end{eqnarray}

On the other hand, choosing the NO-AHD vacuum state defined in the previous section and calling $\tilde{k} = k/\alpha$, one finds the following solution for the function $h_{k}$ \cite{NM}:
\begin{eqnarray} \label{hkh}
        h_{k} = -i\alpha \tilde{k} - 2\alpha x(1-x)\frac{c_{+}c_{-}}{1+i\tilde{k}}\frac{{}_{2}F_{1}\left(c_{+}+1, c_{-}+1, 2+i\tilde{k}; x\right)}{{}_{2}F_{1}\left(c_{+}, c_{-}, 1+i\tilde{k}; x\right)}.
\end{eqnarray}
This function corresponds to the choice $M_{k} = 1/\sqrt{2k}$ and $N_{k} = 0$ in Equation \eqref{ukh_tens}. Using these values and Equation \eqref{NO-AHD modes}, we can evaluate the modes and their derivatives at the end of the bounce period. Then, requiring continuity up to the first time derivative, we obtain initial values for the modes at the beginning of the kinetically dominated epoch.

\subsubsection{Kinetic Domination}

\begin{figure}

    \centering
    \includegraphics[width=16cm]{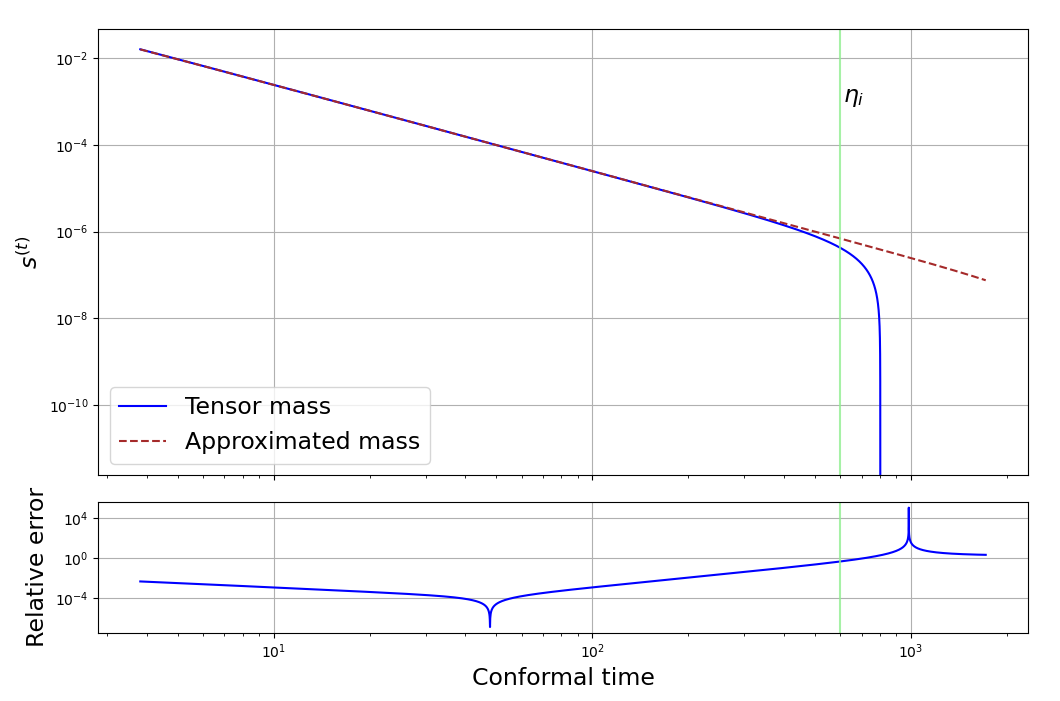}

    \caption{\textbf{Top}: Numerical computation of the tensor mass \eqref{eq_hyb_mass} in kinetic domination in the hybrid approach (solid blue line) and the approximated mass \eqref{eq_Effective_mass_GR} for kinetic domination (dashed red line), both in conformal time. \textbf{Bottom}: Relative error between the numerical and the approximated values of the mass for tensor perturbations in the hybrid approach. We have taken $\gamma=0.2375$ and $\phi_0=0.97$ for the Immirzi parameter and the value of the inflaton at the bounce, respectively, and considered a quadratic inflaton potential with mass $m=1.2 \times 10^{-6}$.}
    \label{fig_Masa_Hyb_T_cinetica}
\end{figure}

This second period can be described using classical relativistic dynamics, because quantum effects are negligible at this stage. The contribution of the inflaton potential is (much) smaller than the kinetic contribution throughout this period, and in this work we will neglect it, as in the previous PT epoch. We show in Figure \ref{fig_Masa_Hyb_T_cinetica} how well the effective mass is estimated by this approximation. If, according to our discussion, we ignore the inflaton potential and solve the background evolution for the free case, we obtain \cite{NM}
\begin{eqnarray} \label{eq_dynamincs_GR}
a(\eta) = a_0 \sqrt{1+2a_0H_0 (\eta-\eta_0)}, \quad \quad \rho(\eta) = \rho_0 \left(\frac{a_0}{a(\eta)}\right)^6, \quad H_0=\sqrt{\frac{8\pi\rho_0}{3}},
\end{eqnarray}
and $\rho_0=\rho(\eta_0)$. Since quantum effects and potential contributions are irrelevant, the effective mass is in practice the same as in GR for the above scale factor, namely
\begin{eqnarray}
    s_\text{GR}^{(t)} (\eta) = \frac{1}{4} \left[(\eta-\eta_0) + \left(\frac{1}{2a_0 H_0}\right) \right]^{-2} . \label{eq_Effective_mass_GR}
\end{eqnarray}

The general solution to the mode equation with this effective mass is \cite{NM}
\begin{eqnarray}
    u_{k}^{KD} (\bar{y}) = \sqrt{\frac{\pi \bar{y}}{4}} \Big[C_k H_0^{(1)} (k\bar{y}) + D_k H_0^{(2)}(k\bar{y}) \Big ],\label{eq_sol_pert_kin}
\end{eqnarray}
where $\bar{y}=\eta-\eta_0 +1/(2a_0H_0) $ and $H_{0}^{(1)}$ and $H_{0}^{(2)}$ are the Hankel functions of the zeroth order and the first and second kinds, respectively. The integration constants $C_{k}$ and $D_{k}$ are fixed by continuity of the modes up to first derivatives, taking as initial conditions the values obtained from the PT solutions at the end of the bounce period (recall that these are in turn determined by the choice of vacuum state). In this manner, we obtain \cite{NM}
\begin{eqnarray}\label{CD}
C_{k} &=&  \frac{1}{H_{0}^{(1)}(\hat{k})} \left[2\sqrt{\frac{k_0}{\pi}}\mu_{k}^{PT}(\eta_0) 
- D_{k} H_{0}^{(2)}(\hat{k})\right],\\
D_{k} &=& \frac{i}{2}\sqrt{\frac{\pi}{k_0}}\left[ k H_{1}^{(1)}(\hat{k}) \mu^{PT}_{k}(\eta_0) 
- \frac{k_0}{2}H_{0}^{(1)}(\hat{k})  \mu^{PT}_{k}(\eta_0)
+ H_{0}^{(1)}(\hat{k}) \dot{\mu}^{PT}_{k}(\eta_0) \right],
\end{eqnarray}
where $\hat{k}=k/k_0$ and $k_0= 2a_0H_0$.

\subsubsection{Slow Roll}

\begin{figure}

    \centering
    \includegraphics[width=16cm]{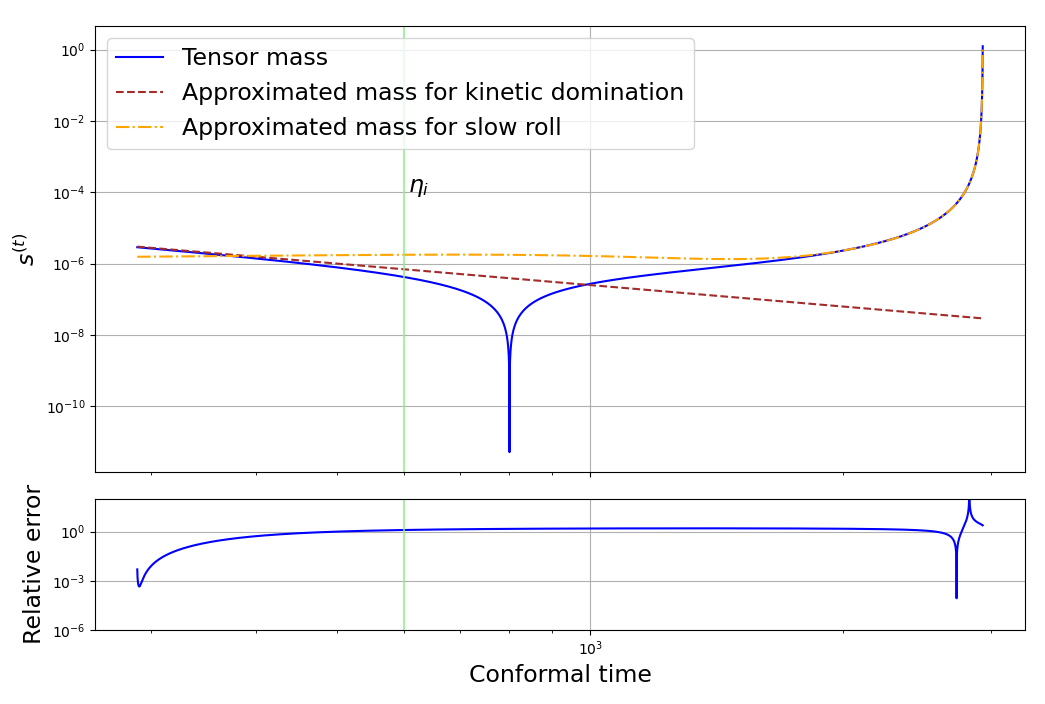}

    \caption{{\textbf{Top}}: Numerical computation of the tensor mass \eqref{eq_hyb_mass} in the hybrid approach (solid blue line), the approximated mass \eqref{eq_Effective_mass_GR} for kinetic domination (dashed red line), and the approximated mass \eqref{eq_def_SR} for slow roll (dash-dotted orange line), all in conformal time. {\textbf{Bottom}}: Relative error between the numerical tensor mass and the approximated tensor mass for slow roll in the hybrid approach. We have taken $\gamma=0.2375$ and $\phi_0=0.97$ for the Immirzi parameter and the value of the inflaton at the bounce, respectively, and considered a quadratic inflaton potential with mass $m=1.2 \times 10^{-6}$.}
    \label{fig_Masa_Hyb_T_SR}
    \end{figure}

The last period considered in our analytic study is the inflationary stage. As the background evolves, the inflaton potential increases and gains relevance in the dynamics (see Figure \ref{fig_Background}). Then, standard slow-roll formulas provide a good approximation, which improves as the potential energy density becomes more and more dominant. We choose the beginning of this period, in the transition from kinetic to potential domination, so that the relative error between the numerical and the approximate effective masses is minimized. This happens at the conformal time $\eta_i \approx 600$, or $t_i \approx 1.88\times 10^4$ in proper time (see Figure \ref{fig_Masa_Hyb_T_SR}).

The slow-roll approximation is based on an expansion of the background equations in terms, e.g., of the parameters
\cite{Langlois, Baumann}
\begin{eqnarray} \label{eq_def_SR}
    \varepsilon_V = \frac{1}{16\pi} \frac{V_{,\phi}^2(\phi)}{V^2(\phi)},  \hspace{2cm} \delta_V = \frac{1}{8\pi} \frac{V_{,\phi\phi}(\phi)}{V(\phi)}. 
\end{eqnarray}
Here, the prime followed by $\phi$ denotes the derivative with respect to the inflaton. During slow-roll inflation, these parameters are approximately constant and much smaller than one, so that a truncation of the expansions at the linear order is sufficient. This leads to the approximation
\begin{equation} \label{eq_prop_SR}
    \frac{d}{d\eta} \left( \frac{a(\eta)}{a'(\eta)} \right) = - (1 - \varepsilon_V).
\end{equation}

We can use Equations \eqref{eq_def_SR} and \eqref{eq_prop_SR}  to derive the expression of the effective mass in slow-roll inflation,
\begin{eqnarray} \label{eq_mass_t_SR}
    s^{(t)}_{SR} = -\frac{\nu^2 - \frac{1}{4}}{(\eta_{e}-\eta)^2}, \hspace{2cm} \nu= \sqrt{\frac{9}{4}+ 3\varepsilon_V},
\end{eqnarray}
where $\eta_{e}$ signals the end of inflation. Using this formula, we can obtain analytically the general solution to the mode equation in the slow-roll period \cite{Langlois,Baumann},
\begin{equation} \label{eq_sol_MS_SR}
    u_{k}^{sr}(\eta) = \sqrt{\frac{\pi}{4}(\eta_{e}-\eta)} \left[A_k H_\nu^{(1)} \left[k(\eta_{e}-\eta)\right] + B_k H_\nu^{(2)} \left[k(\eta_{e}-\eta)\right]\right].
\end{equation}
The Hankel functions are here of the order $\nu$, and the integration constants $A_{k}$ and $B_{k}$ can be determined by imposing continuity of the modes up to their first derivative, matching them to their values at the end of the previous, kinetically dominated era. We obtain \cite{NMY}
\begin{eqnarray}
A_{k} = &&-i\sqrt{\frac{\pi}{16}(\eta_e-\eta_{i})} \Bigg\lbrace kH_{\nu+1}^{(2)}\left[k(\eta_e-\eta_{i})\right] 
- kH_{\nu-1}^{(2)}\left[k(\eta_e-\eta_{i})\right] - \frac{H_{\nu}^{(2)}\left[k(\eta_e-\eta_{i})\right]}{(\eta_e-\eta_{i})} \Bigg\rbrace \mu_{k}^{KD}(\eta_{i}) \nonumber \\
&&
+i\sqrt{\frac{\pi}{4}(\eta_e-\eta_{i})}H_{\nu}^{(2)}\left[k(\eta_e-\eta_{i})\right] \dot{\mu}_{k}^{KD}(\eta_{i}),
\end{eqnarray}
\begin{eqnarray}\label{B}
B_{k} = &&i\sqrt{\frac{\pi}{16}(\eta_e-\eta_{i})} \Bigg\lbrace kH_{\nu+1}^{(1)}\left[k(\eta_e-\eta_{i})\right] 
- kH_{\nu-1}^{(1)}\left[k(\eta_e-\eta_{i})\right] - \frac{H_{\nu}^{(1)}\left[k(\eta_e-\eta_{i})\right]}{(\eta_e-\eta_{i})} \Bigg\rbrace \mu_{k}^{KD}(\eta_{i}) \nonumber \\
&&
-i\sqrt{\frac{\pi}{4}(\eta_e-\eta_{i})}H_{\nu}^{(1)}\left[k(\eta_e-\eta_{i})\right] \dot{\mu}_{k}^{KD}(\eta_{i}).
\end{eqnarray}
With these expressions for the tensor modes, it is straightforward to obtain analytically the~PPS. 

\subsection{Dressed Metric Approach}

We recall that the effective mass of the tensor perturbations in the dressed metric approach, given in Equation \eqref{eq_MS_tens_dress}, is
 \begin{eqnarray} \label{eq_dress_mass}
     \hat{s}^{(t)} = - \frac{4\pi}{3}a^{2}\rho \left(1 + 2\frac{\rho}{\rho_{c}} \right) + 4\pi a^{2}P \left(1 - 2\frac{\rho}{\rho_{c}} \right).
 \end{eqnarray}
The mode equation cannot be solved analytically for this effective mass. To reach an analytic solution, we divide the background evolution into four different periods, in each of which the effective mass can be suitably approximated. First, like in the hybrid approach, we adopt a PT approximation around the bounce. Secondly, and in contrast to the hybrid case, we introduce a matching period with constant effective mass where the quantum effects are already ignorable. From there on, we can estimate the effective mass by its corresponding expression in GR, with two more differentiated relativistic stages like in the hybrid approach, namely, a period of kinetic domination followed by a final period of (slow-roll) inflation.

\subsubsection{Pöschl--Teller Approximation} 

In this period, the contribution of the potential to the effective mass is entirely negligible for the type of background solution that we are analyzing. Recall that the background scale factor takes the form \eqref{eq_a_LQC}. Using this equation, we can compute exactly the effective mass in proper time,
\begin{eqnarray} \label{eq_mass_bounce_dress_analy}
        s^{(t)} = 8\pi a_B^2 \rho_c \frac{8\pi \rho_c (t-t_B)^2-1}{(24\pi\rho_c(t-t_B)^2 -1 )^{5/3}}.
    \end{eqnarray}
However, the relation between the proper time and the conformal time in which the mode equation is expressed is not known analytically, as we saw in the hybrid approach. To circumvent this obstruction, we again use a PT approximation, although with some modifications with respect to the hybrid case. Specifically, we add a constant to the PT potential. This addition allows the effective mass to change its sign, as happens with the exact mass in the evolution. In fact, this last mass is negative at the bounce and becomes positive before the quantum effects dilute, in contrast to the situation found for the hybrid approach, where it is always positive before the onset of inflation. Concretely, we take 
  \begin{equation}\label{eq_PT_dress}
        s_{\text{PT}}^{(t)}(\eta)= \frac{U_0-v_0}{\cosh^2\Big({\alpha(\eta-\eta_\text{B})}\Big)} + v_0.
\end{equation}

The value of the parameters $U_0$, $v_0$, and $\alpha$ can be fixed using the behavior of the exact effective mass. We determine them by imposing that the exact and the approximate effective masses coincide at the bounce, when the exact mass vanishes (which in proper time corresponds to $t_{*} = \sqrt{1/(8\pi \rho_c)} +t_B$), and when this exact mass reaches a maximum (which happens at $t_0= \sqrt{3/(8\pi \rho_c)} +t_B$). This leads to the values $U_0= - 8\pi a_B^2\rho_c$, $v_0= 0.4786$, and $\alpha= 7.915$, the last two of them computed numerically. A numerical analysis also shows that our PT approximation is very satisfactory, with a relative error with respect to the exact effective mass that never exceeds $20\%$ in the whole considered period \cite{AMV}, choosing its final end at $t_0$.

With this approximation, the general solution to the mode equation has the same expression \eqref{ukh_tens} as in the hybrid case, but replacing $k$ with $\bar{k}=\sqrt{ k^2 +v_0}$ and taking different definitions of the variable $x$ and the time-dependent parameters $c_{\pm}$, which are now given by
\cite{AMV}
    \begin{eqnarray}
    x= \left[1+e^{-2\alpha(\eta-\eta_B)}\right]^{-1}, \quad c_{\pm}=\frac{1}{2} \left( 1\pm \sqrt{1+\frac{4\left(U_0-v_0\right)}{\alpha^2}}\right).
    \end{eqnarray}
With these changes, the expression \eqref{hkh}, derived for the function $h_k$ corresponding to the NO-AHD vacuum in the hybrid approach, continues to be valid in the dressed metric case.  Via Equation \eqref{NO-AHD modes}, this function $h_{k}$ determines the mode solution obtained from Equation~\eqref{ukh_tens} by setting $M_{k} = 1/\sqrt{2\bar{k}}$ and $N_{k} = 0$, formally as in the hybrid approach.
    
\subsubsection{Constant Effective Mass} 

In order to better approximate the behavior of the effective mass, it was shown in Ref. \cite{AMV} that it is convenient to introduce a transition period between the bounce and the kinetic regime. In this period, the effective mass can be considered to be constant and equal to its value $m_0$ at the end of the bounce period,
    \begin{eqnarray} \label{eq_Effective_mass_cste}
    s_{0}^{(t)} (\eta) = m_0= \frac{U_0-v_0}{\cosh^2\Big({\alpha(\eta_{0}-\eta_\text{B})}\Big)} + v_0  ,
    \end{eqnarray}
where $\eta_0$ is the conformal time corresponding to $t_0$.

\begin{figure}

\centering
    \includegraphics[width=16cm]{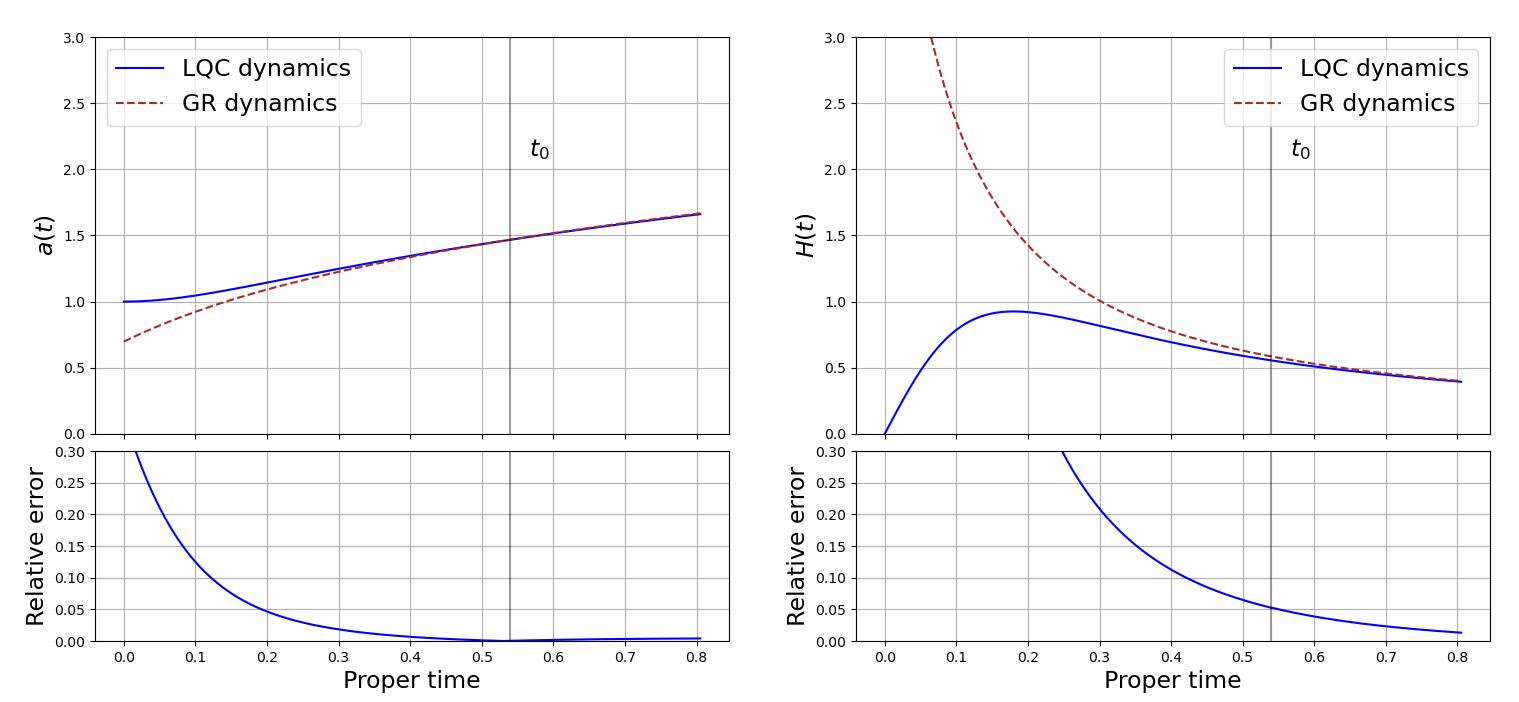}

    \caption{\textbf{Top}: Numerical computation in effective LQC (solid blue line) and in GR (dashed red line) of the background scale factor (\textbf{left}) and the Hubble parameter (\textbf{right}) in kinetic domination, given by Equations \eqref{eq_a_LQC} and \eqref{eq_dynamincs_GR}. Here, $t_0=\sqrt{3/(8\pi\rho_c)} + t_B $. \textbf{Bottom}: Relative error between the two scale factors and the two Hubble parameters. We have taken $\gamma=0.2375$ and $\phi_0=0.97$ for the Immirzi parameter and the value of the inflaton at the bounce, respectively, and considered a quadratic inflaton potential with mass $m=1.2 \times 10^{-6}$.}
    \label{fig_a_vs_LQC_GR_2}
    \end{figure}

The LQC corrections to the background are already ignorable and the background evolves as in standard relativistic cosmology during a kinetically dominated era, as one can see in Figure \ref{fig_a_vs_LQC_GR_2}. The general solution to the mode equation is  
\begin{eqnarray}
    u_k(\eta) = \alpha_k e^{i\kappa\eta  } + \beta_k  e^{-i\kappa\eta},
\end{eqnarray}
where $\kappa=\sqrt{k^2 + m_{0}}$. To fix the constants $\alpha_k$ and $\beta_k$, we use the continuity of the mode up to its first derivative in the matching with the PT epoch. In this way, we obtain
\begin{eqnarray}
\alpha_k=   \frac{e^{-i\kappa\eta_0}}{2\kappa}\left[\kappa u_k(\eta_0) -i u_k'(\eta_0)\right] ,\quad
\beta_k= \frac{ e^{i\kappa\eta_0}}{2\kappa}\left[\kappa u_k(\eta_0)+i u_k'(\eta_0)\right]. \label{eq::beta_k_cste}
\end{eqnarray}

\subsubsection{Kinetic Domination} 

\begin{figure}

\centering 
    \includegraphics[width=16cm]{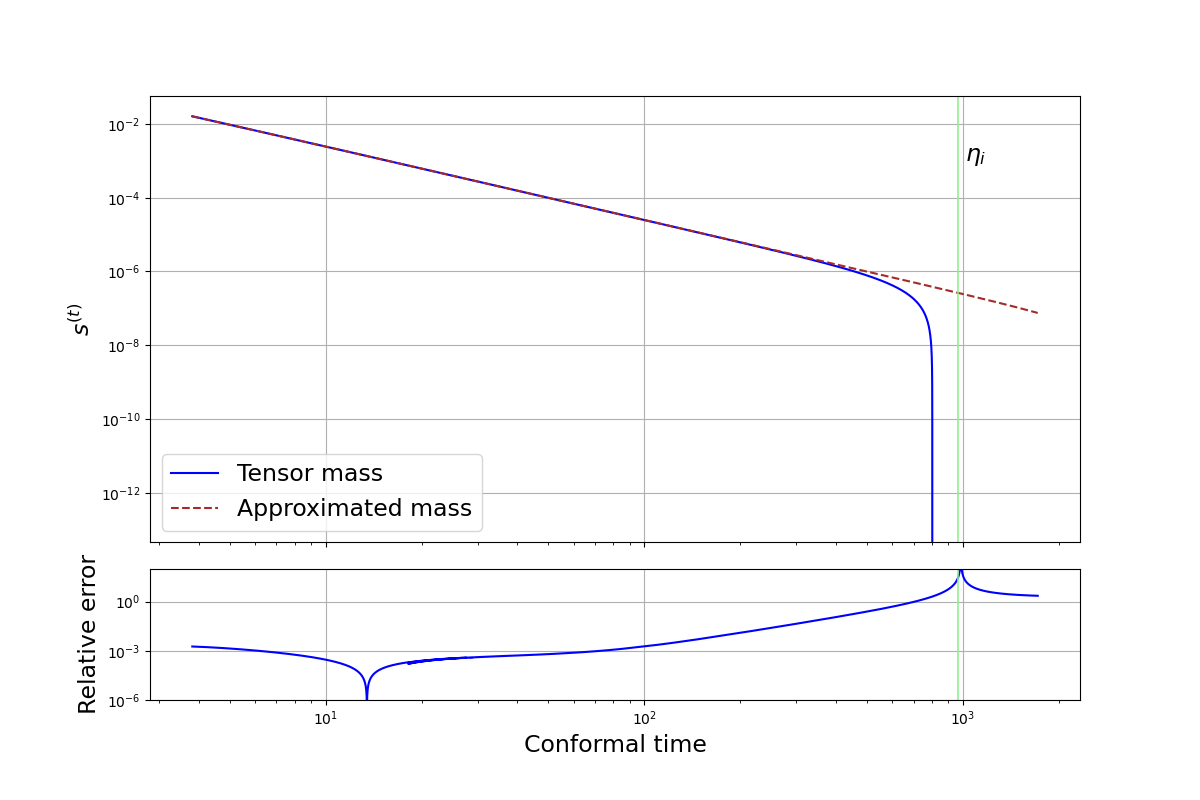}
    \caption{\textbf{Top}: Numerical computation of the tensor mass \eqref{eq_dress_mass} in the dressed metric approach (solid blue line) and the approximated mass \eqref{eq_Effective_mass_GR} for kinetic domination (dashed red line), both in conformal time. \textbf{Bottom}: Relative error between the numerical and the approximated values of the mass for tensor perturbations in the dressed approach. We have taken $\gamma=0.2375$ and $\phi_0=0.97$ for the Immirzi parameter and the value of the inflaton at the bounce, respectively, and considered a quadratic inflaton potential with mass $m=1.2 \times 10^{-6}$.}
    \label{fig_Masa_Dress_T_cinetica}
    \end{figure}

In this period, quantum effects are also irrelevant, and the contribution of the potential to the inflaton energy density can be ignored. Then, the effective mass again takes the form \eqref{eq_Effective_mass_GR}. We compare this approximation with the exact effective mass in Figure \ref{fig_Masa_Dress_T_cinetica}. We fix the beginning of this period at the moment when $s_\text{GR}^{(t)} (\eta_0') = m_0$, which is 
\begin{eqnarray}
    \eta_0'= \eta_0 + \frac{1}{\sqrt{4m_0}} - \left(\frac{1}{2a_0H_0}\right).
\end{eqnarray}
     
The general solution to the mode equation with our approximation is the same as in Equation \eqref{eq_sol_pert_kin}. 
However, the constants $C_k$ and $D_k$ now take different values. To determine them, we use the continuity of the perturbations up to the first time derivative at the matching point with the period of constant effective mass. This condition leads to
\begin{eqnarray}
\notag C_k &=& \frac{ i }{4} \Bigg[\alpha_k e^{i\kappa\eta_0'} \left( \sqrt{\frac{\pi}{\bar{y}_0'}}\left(1 -2i\kappa \bar{y}_0'\right) H_0^{(2)}(k\bar{y}_0') - 2\sqrt{\pi \bar{y}_0' k^2} H_1^{(2)}(k\bar{y}_0')   \right) \\&+& \beta_k e^{-i\kappa\eta_0'} \left(  \sqrt{\frac{\pi}{\bar{y}_0'}}\left(1 +2i\kappa \bar{y}_0'\right) H_0^{(2)}(k\bar{y}_0') - 2\sqrt{\pi \bar{y}_0' k^2} H_1^{(2)}(k\bar{y}_0')   \right) \Bigg]\label{eq::C_cste},\\
\notag D_k &=&  -\frac{ i }{4}\Bigg[ \alpha_k e^{i\kappa\eta_0'}\left( \sqrt{\frac{\pi}{\bar{y}_0'}} \left(1 - 2i \kappa\bar{y}_0'\right) H_0^{(1)}(k\bar{y}_0') - 2\sqrt{\pi \bar{y}_0' k^2} H_1^{(1)}(k\bar{y}_0')  \right) \\&+&  \beta_k e^{-i\kappa\eta_0'} \left(   \sqrt{\frac{\pi}{\bar{y}_0'}} \left(1 +2 i \kappa\bar{y}_0'\right) H_0^{(1)}(k\bar{y}_0') - 2\sqrt{\pi \bar{y}_0' k^2} H_1^{(1)}(k\bar{y}_0') \right)\Bigg] ,   \label{eq::D_cste}
\end{eqnarray}
where $\bar{y}_0'=\eta_0' - \eta_0 + 1/(2a_\text{0}H_0) $, and $\kappa$, $\alpha_k$, and $\beta_k$ were given in the previous subsection.

\subsubsection{Slow Roll} 

In this last period, we finally reach an inflationary regime, which (at least during most of it) admits a slow-roll description. The effective masses of the hybrid and dressed metric approaches become almost identical, as is realized in Figure \ref{fig_difference}. Therefore, we use the same type of approximation employed in the hybrid case. 

\section{Primordial Power Spectrum}\label{sec6}

We can now proceed to compute the PPS, both by analytic means, using the different approximations to the mode solutions that we have presented, and by numerical methods, employing the exact expression of the effective mass and integrating the mode equation. We do so for the two quantization approaches within LQC studied in this paper.

The approximations that we have introduced and justified provide the necessary ingredients to calculate the PPS of the tensor perturbations analytically. In doing so, we do not need to evaluate the modes at the end of inflation, but rather when they are all frozen (at a time $\eta_{fr}$). In the case of our numerical computations, this significantly reduces the calculation time. Moreover, it supports the use of slow-roll approximation during inflation if it is valid until $\eta_{fr}$. Then, the PPS is defined by the expression \cite{Langlois,Baumann}
    \begin{eqnarray}\label{eq__def_PPS_T}
        \mathcal{P}_\text{T} = \frac{32k^{3}}{\pi} \frac{|u_k(\eta_{fr})|^2}{a(\eta_{fr})^2}.
    \end{eqnarray}

For all wavenumbers $k$ in the relevant window for observations, the argument of the Hankel functions in our analytic formula \eqref{eq_sol_MS_SR} for the solutions during slow-roll inflation turns out to be much smaller than one. Hence, we can use the properties of these Hankel functions to approximate them \cite{Abra}, writing the solution $u_k^{sr}$ as 
    \begin{eqnarray}
        |u_k^{sr}| \simeq \frac{1}{4\pi} (\eta_{e}-\eta_{fr}) |\Gamma(\nu)|^2 \left[\frac{k(\eta_{e}-\eta_{fr})}{2}\right]^{-2\nu} |A_k - B_k|^2,
    \end{eqnarray}
where $\Gamma$ is the gamma function. The corresponding expression of the PPS reads
    \begin{eqnarray} \label{eq_C_v_T}
        \mathcal{P}_\text{T} (k) \simeq C_\nu k^{3-2\nu} |A_k-B_k|^2, \quad \text{where} \quad C_\nu = \frac{16 |\Gamma(\nu)|^2  }{\pi^2 a_{fr}^2} \left(\frac{\eta_{e}-\eta_{fr}}{2}\right)^{1-2\nu}.
    \end{eqnarray}
Here, $a_{fr}$ is the value of the scale factor at $\eta_{fr}$. 

Note that the PPS critically depends on the integration constants $A_k$ and $B_k$, which determine the mode solutions for the perturbations. Normalization of these solutions only imposes the condition $u_k u^{*\prime}_k-u^{\prime}_ku^{*}_k=1$, which amounts to the restriction $|B_k|^2-|A_k|^2=1$ on the norms of $A_k$ and $B_k$. The symbol $*$ stands for complex conjugation. We therefore see that the PPS depends on the choice of mode solutions specified by the vacuum state, and different choices yield different spectra. The form of the spectrum in the above expression leads to fast oscillations in $k$ because, even if the norms of $A_k$ and $B_k$ vary slowly, their phases $\theta_k^A$ and $\theta_k^B$ often give rise to a rapidly oscillating interference. Concretely,  
\begin{equation}\label{PPSosci}
\mathcal{P}_\text{T}(k) = C_\nu k^{3-2\nu} \Big[|A_k|^2 + |B_k|^2 - 2 |A_k||B_k| \cos{(\theta_k^A - \theta_k^B)}\Big].
\end{equation}
Nonetheless, it is possible to remove these undesired oscillations in $k$, which could artificially pump power into the spectrum on average. Indeed, as was argued and supported in Ref. \cite{NM}, we can adjust the choice of mode solutions using the following Bogoliubov transformation that leads to a state really free of spurious oscillations in the PPS: 
 \begin{eqnarray}
A_k \to \Tilde{A}_k = |A_k|, \quad \quad B_k \to \Tilde{B}_k = |B_k|.
\end{eqnarray}
The corresponding PPS is just
\begin{eqnarray}
\mathcal{P}_\text{T}(k) = C_\nu k^{3-2\nu} \left(|A_k| - |B_k|\right)^2 .\label{eq_PW_formula_cstes}
\end{eqnarray}
We notice that the spectrum obtained with this Bogoliubov transformation is the envelope of the minima of the previous one in Equation \eqref{PPSosci}. This new PPS is free from rapid oscillations, but retains all other relevant information about the scale dependence of the original spectrum. 

On the other hand, we also want to compute the PPS of the NO-AHD vacuum state in a numerical way, a computation that has not been performed before in the literature. This will allow us to test the goodness of our analytic approach, checking its agreement with the exact numerical result. In order to numerically solve the dynamics of the perturbations, we impose initial conditions at the time $\eta_0$ that marks the end of the bounce period and integrate from there until the instant when the mode freezes during inflation. We perform the numerical integration with an Adams--Bashforth--Moulton method and a Runge--Kutta method of the eighth order, obtaining the same result in both cases. Since the PT potential approximates well the effective mass during the bounce period and allows us to determine an NO-AHD state for the tensor modes, we use as our initial conditions the values of the modes and their time derivatives at $\eta_0$ obtained with this choice of state and with that approximation. In particular, using Equations \eqref{NO-AHD modes} and \eqref{Riccati}, we find that the time derivative is given by
\begin{eqnarray}
        \dot{u}_k (\eta_0) = - h_k^*(\eta_0) \frac{u_k(\eta_0)}{a(\eta_0)}.
\end{eqnarray}

\begin{figure}

                \includegraphics[width=14cm]{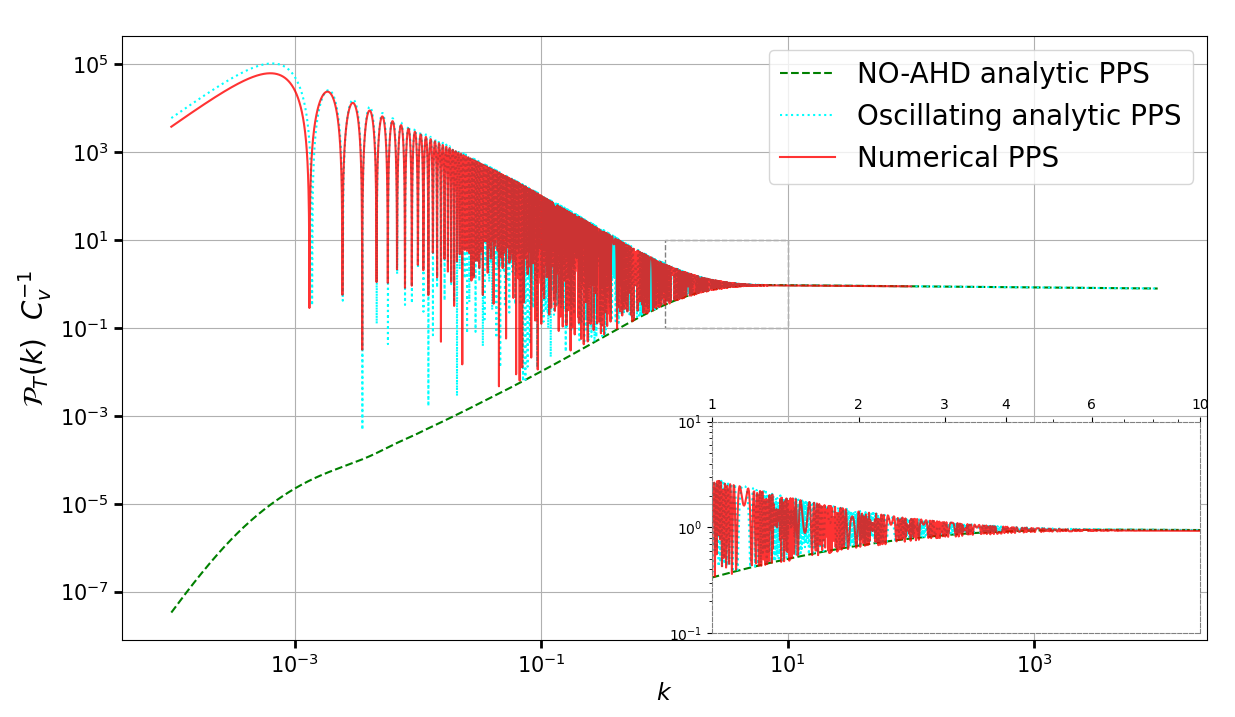}
                \caption{Normalized primordial power spectrum (PPS) for the hybrid approach. We show the numerical computation of the oscillating PPS (solid red line), the analytic approximation to this PPS  (dotted cyan line), and the analytic approximation to the PPS of the NO-AHD vacuum (dashed green line), obtained with a Bogoliubov transformation. We include an inset enlarging the region $1\leq k \leq 10$ (framed in the PPS) to see the details when power suppression appears. We have taken $\gamma=0.2375$ and $\phi_0=0.97$ for the Immirzi parameter and the value of the inflaton at the bounce, respectively, and considered a quadratic inflaton potential with mass $m=1.2 \times 10^{-6}$.}
                \label{fig_PPS_T_Hyb_Norma}
\end{figure}

\begin{figure}

\centering 
                \includegraphics[width=15cm]{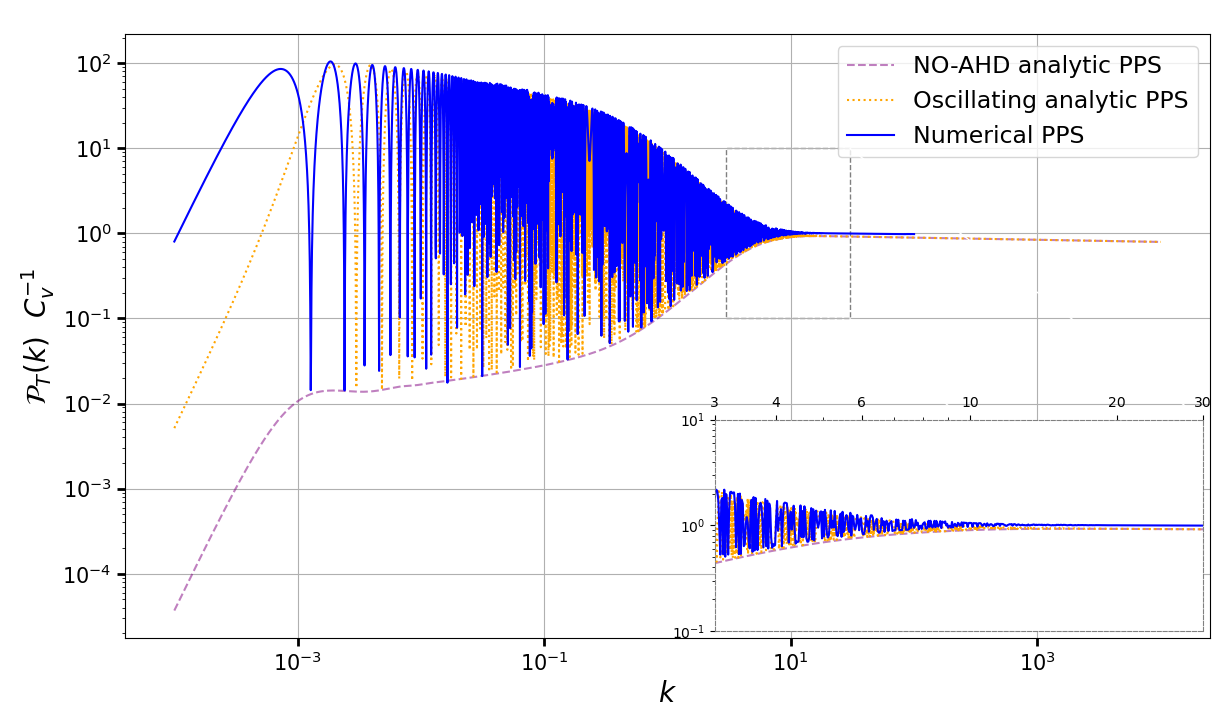}

                \caption{Normalized primordial power spectrum (PPS) for the dressed metric approach. We show the numerical computation of the oscillating PPS (solid blue line), the analytic approximation to this PPS (dotted orange line), and the analytic approximation to the PPS of the NO-AHD vacuum (dashed violet line), obtained with a Bogoliubov transformation. We include an inset enlarging the region $3\leq k \leq 30$ (framed in the PPS) to see the details when power suppression appears. We have taken $\gamma=0.2375$ and $\phi_0=0.97$ for the Immirzi parameter and the value of the inflaton at the bounce, respectively, and considered a quadratic inflaton potential with mass $m=1.2 \times 10^{-6}$.}
                \label{fig_PPS_T_DS_Norma}
        \end{figure}

After numerical integration of the modes, the PPS is calculated using definition \eqref{eq__def_PPS_T}. For a normalized spectrum, we divide this expression by $C_\nu$, defined in Equation \eqref{eq_C_v_T}. In Figure \ref{fig_PPS_T_Hyb_Norma}, we display our analytic approximation to the PPS for the hybrid approach, with and without the final Bogoliubov transformation that leads to the desired non-oscillating state. We also plot the PPS of the hybrid case calculated by means of our numerical integration, without the final Bogoliubov transformation. We see that the oscillatory analytic spectrum reproduces very well its numerical counterpart. Moreover, the non-oscillatory PPS corresponds extremely well to the envelope of the minima of the oscillatory ones, not only of the analytic computation but also of the numerical spectrum. This important result, which strongly supports our analytic description of the non-oscillating PPS, is also reached in the case of the dressed metric approach, as we can check from the plot of the corresponding spectra shown in Figure \ref{fig_PPS_T_DS_Norma}. In this last case, nonetheless, our graphics seem to indicate a possible slight difference in the frequency of the oscillations between the analytic and the numerical spectra.

\begin{figure}

\centering 
    \includegraphics[width=16cm]{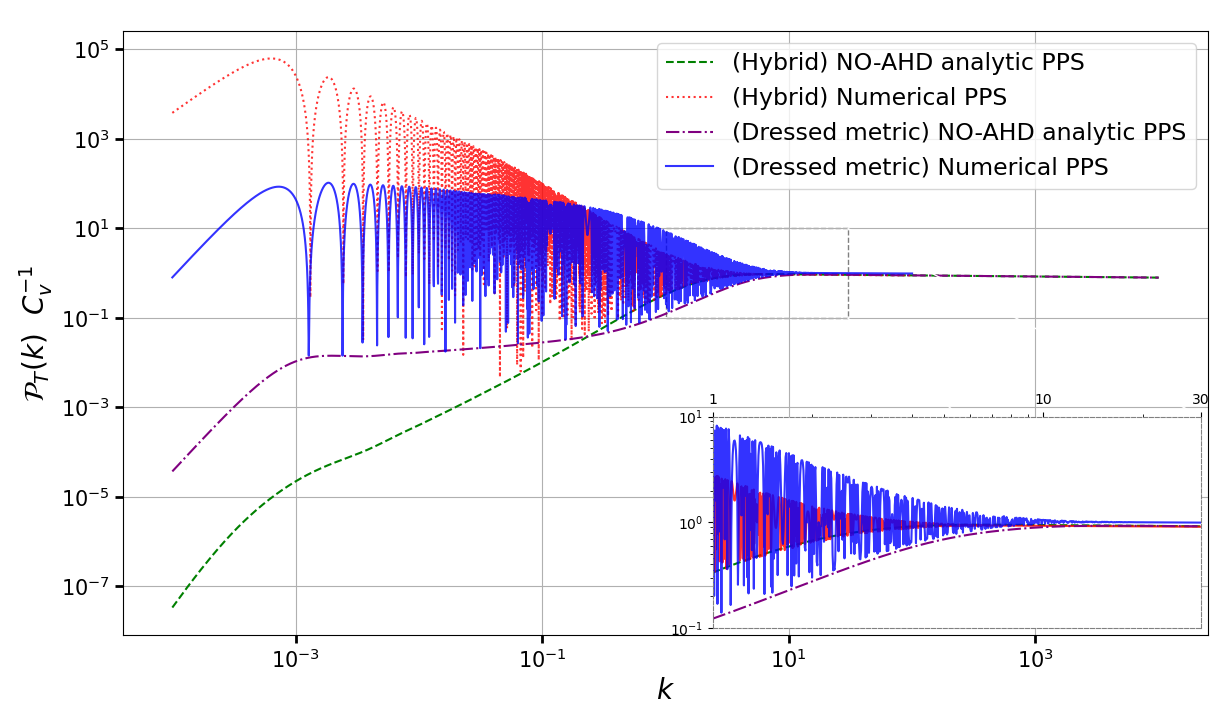}

    \caption{Comparison of the normalized primordial power spectra (PPS) for the hybrid and the dressed metric approaches. We show the numerical computation of the oscillating PPS for the hybrid approach (dotted red line) and for the dressed metric approach (solid blue line), as well as the analytic approximations to the PPS of the NO-AHD vacuum for the hybrid approach (dashed green line) and the dressed metric approach (dash-dotted violet line). We include an inset enlarging the region $1\leq k \leq 30$ (framed in the PPS) to see the details when power suppression appears. We have taken $\gamma=0.2375$ and $\phi_0=0.97$ for the Immirzi parameter and the value of the inflaton at the bounce, respectively, and considered a quadratic inflaton potential with mass $m=1.2 \times 10^{-6}$.}
    \label{fig_PPS_junts}
    \end{figure}

We can also compare the difference between the spectra of the two considered quantization approaches. Possibly, the most remarkable fact is that the power suppression in infrared scales is clearly greater for the hybrid approach, as we can see in Figure \ref{fig_PPS_junts}. Furthermore, the decrease in power around the $k$-cutoff where the suppression begins is more pronounced in the PPS of the hybrid approach. This cutoff, although similar, is not exactly the same for the two approaches. It seems a bit larger in the dressed metric case. Nevertheless, the spectra of both approaches become almost identical in the quasi-invariant region, corresponding to modes beyond the cutoffs. In this region, we observe the tilt introduced by slow-roll inflation. 

To conclude this section, let us comment that the window of observable modes corresponds, approximately, to the interval $ 2 \times10^{-4} {\rm Mpc}^{-1} \leq k/a_{today} \leq 6 \times 10^{-1} {\rm Mpc}^{-1}$, which therefore depends on the background cosmology via the present value of the scale factor, $a_{today}$ \cite{Planck}. Since an inverse megaparsec is approximately $5 \times 10^{-58}$ inverse Planck lengths, and $a_{today}=e^{n_T}$ with our convention that $a_{B}=1$, where $n_T$ denotes the number of e-folds from the bounce until nowadays, it is easy to conclude that the observable window is approximately $[1,3\times 10^3]\times10^{-61} e^{n_T}$. In particular, this whole window includes wavenumbers in the interval $[10^{-1}, 10]$ around the cutoff if, roughly speaking, $130 \leq n_T \leq 143$. We note that a number of total e-folds of this order has been obtained in numerical calculations for the background carried out in previous works, for instance in Refs. \cite{AG1,Morris}.

\vspace{-12pt}

\section{Discussion}\label{sec7}

We have investigated the implications of two different approaches to the quantization of cosmological tensor perturbations within LQC, focusing our discussion on their PPS. More specifically, we have considered the hybrid and the dressed metric approaches. Among the various approaches developed in LQC, these two have the nice property that they do not affect the hyperbolicity of the field equations for the perturbations in the ultraviolet, nor do they modify the dispersion relations for the perturbation modes in this sector. For completeness in our discussion, we have presented a brief overview of these approaches, pointing out their similitudes and differences. The hybrid approach rests on a canonical quantization of the entire constrained system formed by the background geometry and its perturbations, described by the action truncated at the quadratic perturbative order. In contrast, the dressed metric approach first quantizes the background geometry, incorporates the most important quantum effects resulting in a background with a dressed metric, and then lifts its effective trajectories to the unconstrained phase space of the perturbations, treating such perturbations as test fields propagating on the dressed background. The fact that the classical relation between canonical momenta and time derivatives of the background metric is not respected in effective LQC implies that the outcome of the two approaches differs when one considers quantum background states peaked on effective trajectories.  

In both approaches, the field equations of the perturbations can be expressed as generalized wave equations, with a background-dependent effective mass. We have provided the expression of this effective mass in both approaches for perturbations of the tensor type. The corresponding effective masses are 
almost equal soon after the bounce experienced in (effective) LQC, around a fraction of a Planck second after it. In fact, from this moment on, the quantum effects on the background and on the effective mass of the perturbations are negligible. However, around the bounce, the two considered effective masses are very different, leading to distinct behaviors both in the evolution of tensor modes and in the vacuum state that is optimally adapted to the background. These differences are transmitted to the PPS. Actually, the effective mass for the dressed metric approach is negative at the bounce and flips its sign later in the evolution, while the effective mass of the hybrid approach turns out to be positive in the whole interval from the bounce until the onset of inflation for the effective background solutions that we have studied in LQC.

The propagation of the (Fourier) tensor modes cannot be resolved analytically in any of the two approaches. To circumvent this problem, we have divided the evolution into several eras where the effective mass can be approximated conveniently so as to render the dynamics of the tensor perturbations solvable.  In the hybrid approach, we have considered three different evolution periods, whereas for the dressed metric we have included an additional period to improve the approximation in the transition from the quantum to the classical regime, around the time when the effective mass displays a maximum (soon after it vanishes). In the last period describing the inflationary expansion, we have used a slow-roll approximation to consider the influence of the (not exactly constant) inflaton potential. We have motivated all these approximations and checked numerically that the relative error that they introduce in the effective mass is acceptable.

Since the considered background includes a preinflationary epoch in which the background is nonstationary and which can significantly affect modes contained in the observable window, in principle there is no preferred state that can serve as a natural vacuum to construct a quantum field theory for (these modes of) the perturbations. Nevertheless, several proposals have been put forward for a preferred choice of vacuum in this scenario. In this work, we have adhered to the NO-AHD proposal. This proposal selects a state optimally adapted to the background (at least in the asymptotic ultraviolet sector), inasmuch as it can be specified by an asymptotic diagonalization of the Hamiltonian that governs the dynamics of the tensor perturbations, and leads to slowly varying mode amplitudes that provide a non-oscillating PPS \cite{NMT,NM}. In general, rapid variations can be interpreted as a poor adaptation to the background, resulting in a transfer from the background evolution to changes in the mode amplitudes. Associated high-frequency oscillations typically pump spurious power to the PPS. Actually, the choice of an NO-AHD vacuum state has already been studied in the literature for the two analyzed quantization approaches \cite{NM,NMY,AMV}. Although the exact determination of such a state by analytic means is not completely possible, a suitable approximation of the effective mass in the quantum region around the bounce allows us to obtain its expression. This expression is different in the two approaches, reflecting the differences in the corresponding effective masses at the bounce. We have used the corresponding values of the tensor modes and their time derivatives at the end of the bounce period as our initial conditions for the integration of the propagation of the perturbations in the rest of the time interval until the end of inflation. The approximation used in this procedure (together with numerical errors or further analytic approximations) still introduces a fast-oscillating phase in the integration of the modes, but this phase is harmless for our purposes, because it can be absorbed by a suitable adjustment of our state, implemented as a Bogolibouv transformation. As we have shown, this transformation finally provides us with the desired non-oscillating PPS for the tensor perturbations.  

With all these ingredients, we have calculated the PPS for both quantization approaches. In addition to our approximate analytic derivation, we have presented a procedure to numerically compute this tensor sprectrum for the two approaches for the first time in the literature. The main result of our study is that the analytic approximation fits remarkably well the exact PPS, obtained with numerical methods. This exact non-oscillating PPS is the envelope of the minima of the non-adjusted oscillating spectrum. This result is equally valid for the hybrid and the dressed metric approaches. In addition, comparing the spectra of the two quantization approaches, we have seen that the hybrid proposal leads to a larger power suppression in infrared scales. Moreover, there is a cutoff scale in both approaches of the order of the physical scale corresponding, roughly speaking, to the Planck curvature scale at the bounce. This cutoff scale is similar in both cases, although it is a little bigger in the dressed metric approach. In addition, the decrease in power around the corresponding cutoff scale is steeper in the hybrid spectrum. Our results are robust, both in the sense that we obtain the same spectra using analytic approximations and numerical techniques, and inasmuch as the cutoff scales are of comparable order in the two studied approaches.

These results are extremely helpful to simplify the computations for a future parametrization of the PPS in LQC.  They strongly support the validity of our approximations, justifying the use of analytic expressions in which the dependence on the parameters of the model can be handled explicitly without the need for highly demanding numerical computations. It would be enlightening to consider distinct approximations in the bounce region that improve the effective mass below our maximum relative error during this epoch preceding the time interval of our numerical calculations (we recall that this maximum is approximately $25\%$ or $20\%$, respectively, in the hybrid and dressed metric approaches). Nevertheless, we do not anticipate radical changes in the PPS of the NO-AHD vacuum, given that the bounce epoch lasts for a fraction of a Planck second and our numerical computations already show that this PPS is not extremely sensitive to variations in the effective mass in this period. In this respect, note that the effective mass differs considerably in the hybrid and dressed metric approaches close to the bounce, as shown in the top part of Figure~\ref{fig_difference}, but the respective PPS for these approaches have comparable cutoff scales and power suppression near them. 

Other possible questions for further research concern the numerical computation of the B-mode polarization power spectrum and a detailed study of the quantum geometry effects on the CBGW. A parametrization of the PPS based on our results will facilitate Bayesian analyses to discuss the dependence of the spectra not only on the parameters of our approximated description, including the part characterizing the inflaton potential, but also on the standard cosmological parameters that affect the evolution of the perturbations after the end of inflation. For instance, extrapolating observational data from scalar to the tensor perturbations here discussed, we have argued that the observable window of modes can overlap with the scale of power suppression that we have found, provided that the total number of e-folds is, roughly speaking, between 130 and 143. Accepting some 70 e-folds from the bounce until the end of inflation, an estimation compatible with our calculations (see Figure \ref{fig_Background}), we conclude that the background should experience approximately 60 to 73 e-folds after inflation, with the upper end of this interval favoring scales of power suppression in the lower end of the observable window, and therefore corresponding to the optimal case for a lack of power at low multipoles in the angular power spectrum of the CMB. It would be very interesting to explore the regions of the parameter space where this happens. These studies would provide an important input for understanding the Very Early Universe. 

\acknowledgments

This work was supported by grants PID2020-118159GB-C41, PID2022-138626NB-I00, RED2022-134204-E, and RED2022-134411-T, funded by MCIN/AEI/10.13039/501100011033/FEDER, UE; the Universitat de les Illes Balears (UIB); the MCIN with funding from the European Union NextGenerationEU/PRTR (PRTR-C17.I1); the Comunitat Autonòma de les Illes Balears through the Direcció General de Recerca, Innovació I Transformació Digital with funds from the Tourist Stay Tax Law (PDR2020/11-ITS2017-006), the Conselleria d’Economia, Hisenda i Innovació grants No. SINCO2022/18146 and SINCO2022/6719, co-financed by EU and FEDER Operational Program 2021-2027 of the Balearic Islands; and the “ERDF A way of making Europe”. J.Y.C. is supported by the Spanish MICIU via an FPI doctoral grant (PRE2022-000809).The authors are thankful to B. Elizaga Navascu\'es for important contributions in the development of this work. They are also thankful to A. Alonso-Serrano and P. Santo-Tomás for conversations.


\begin{thebibliography}{299}

\bibitem{LQG} Ashtekar, A.; Lewandowski, J. Background independent quantum gravity: A Status report. {\em Class. Quant. Grav.} {\bf 2004}, {\em 21}, R01.

\bibitem{Thie} Thiemann, T. {\em Modern Canonical Quantum General Relativity}; Cambridge University Press: Cambridge, UK, 2007.

\bibitem{LQC} Ashtekar, A.; Singh, P. Loop quantum cosmology: A status report. {\em Class. Quant. Grav.} {\bf 2011}, {\em 28}, 213001.

\bibitem{APS} Ashtekar, A.; Paw{\l}owski, T.; Singh, P. Quantum nature of the big bang. {\em Phys. Rev. Lett.} {\bf 2006}, {\em 96}, 141301.

\bibitem{APS1} Ashtekar, A.; Paw\l{}owski, T.; Singh, P. Quantum nature of the big bang: Improved dynamics. {\em Phys. Rev. D} {\bf 2006}, {\em 74}, 084003.

\bibitem{Planck} Aghanim, N. { et al.} [Planck Collaboration]. Planck 2018 results. VI. Cosmological parameters. {\em Astron. Astrophys.} {\bf 2020}, {\em 641}, A6; Erratum in {\em Astron. Astrophys.} {\bf 2021}, {\em 652}, C4]. 

\bibitem{CBGW} Caprini, C.; Figueroa, D.G. Cosmological backgrounds of gravitational waves. {\em Class. Quant. Grav.} {\bf 2018}, {\em 35}, 163001.

\bibitem{LISA} Auclair, P. {et al.} [LISA Collaboration]. Cosmology with the Laser Interferometer Space Antenna. {\em Living Rev. Rel.} {\bf 2023}, {\em 26}, 5.

\bibitem{NANOGrav} Agazie, G. {et al.} [NANOGrav Collaboration]. The NANOGrav 15 yr data set: Evidence for a gravitational-wave background. {\em Astrophys. J. Lett.} {\bf 2023}, {\em 951}, L8.

\bibitem{NANOGrav2} Afzal, A. { et al.} [NANOGrav Collaboration]. The NANOGrav 15 yr data set: Search for signals from new physics. {\em Astrophys. J. Lett.} {\bf 2023}, {\em 951}, L11.

\bibitem{ASr} Ashtekar, A.; Gupt, B.; Jeong, D.; Sreenath, V. Alleviating the tension in the cosmic microwave background using Planck-scale physics. {\em Phys. Rev. Lett.} {\bf 2020}, {\em 125}, 051302.

\bibitem{ASr2} Ashtekar, A.; Gupt, B.; Sreenath, V. Cosmic tango between the very small and the very large: Addressing CMB anomalies through loop quantum cosmology. {\em Front. Astron. Space Sci.} {\bf 2021}, {\em 8}, 685288.

\bibitem{AgSr} Agullo, I.; Kranas, D.; Sreenath, V. Large scale anomalies in the CMB and non-Gaussianity in bouncing cosmologies. {\em Class. Quant. Grav.} {\bf 2021}, {\em 38}, 065010.

\bibitem{AgSr2} Agullo, I.; Kranas, D.; Sreenath, V. Anomalies in the cosmic microwave background and their non-Gaussian origin in loop quantum cosmology. {\em Front. Astron. Space Sci.} {\bf 2021}, {\em 8}, 703845.

\bibitem{Paper_2} Mena Marug\'an, G.A.; Vicente-Becerril, A.; Y\'ebana Carrilero, J. Analytic and numerical power spectra of scalar perturbations in loop quantum cosmology. {\em Phys. Rev. D} {\bf 2024}, {\em 110}, 043508.

\bibitem{Mukhanov1} Mukhanov, V. {\em Physical Foundations of Cosmology}; Cambridge University Press: Cambridge, UK, 2005.

\bibitem{Bunch} Bunch, T.S.; Davies, P.C.W. Quantum field theory in de Sitter space: Renormalization by point splitting. {\em Proc. R. Soc. Lond. A} {\bf 1978}, {\em 360}, 117--134.

\bibitem{Baumann} Baumann, D. Inflation. In {\em Physics of the Large and the Small, TASI 2009};  Csaki, C., Dodelson, S., Eds.;  World Scientific: Singapore, 2011; pp. 523--686.  

\bibitem{Langlois} Langlois, D. Inflation and cosmological perturbations. {\em Lect. Notes Phys.} {\bf 2010}, {\em 800}, 1. 

\bibitem{hybr_rev} Elizaga Navascu\'es, B.; Mena Marug\'an, G.A. Hybrid loop quantum cosmology: An overview. {\em Front. Astron. Space Sci.} {\bf 2021}, {\em 8}, 624824.

\bibitem{dressed1} Agullo, I.; Ashtekar, A.; Nelson, W. A quantum gravity extension of the inflationary scenario. {\em Phys. Rev. Lett.} {\bf 2012}, {\em 109}, 251301.

\bibitem{dressed2} Agullo, I.; Ashtekar, A.; Nelson, W. Extension of the quantum theory of cosmological perturbations to the Planck era. {\em Phys. Rev. D} {\bf 2013}, {\em 87}, 043507.

\bibitem{effective2} Bojowald, M.; Calcagni, G.; Tsujikawa, S. Observational constraints on loop quantum cosmology. {\em Phys. Rev. Lett.} {\bf 2011}, {\em 107}, 211302.

\bibitem{effective3} Cailleteau, T.; Linsefors, L.; Barreau, A. Anomaly-free perturbations with inverse-volume and holonomy corrections in loop quantum cosmology. {\em Class. Quant. Grav.} {\bf 2014}, {\em 31}, 125011.

\bibitem{effective4} Wilson-Ewing, E. Holonomy corrections in the effective equations for scalar mode perturbations in loop quantum cosmology. {\em Class. Quant. Grav.} {\bf 2012}, {\em 29}, 085005.

\bibitem{effective5} Bolliet, B.; Grain, J.; Stahl, C.; Linsefors, L.; Barrau, A. Comparison of primordial tensor power spectra from the deformed algebra and dressed metric approaches in loop quantum cosmology. {\em Phys. Rev. D} {\bf 2015}, {\em 91}, 084035.

\bibitem{hybr_ref} Castell\'o Gomar, L.; Mart\'{\i}n-Benito, M.; Mena Marug\'an, G.A. Gauge-invariant perturbations in hybrid quantum cosmology. \emph{JCAP J. Cosmol. Astropart. Phys.} {\bf 2015}, {\em 6}, 045.

\bibitem{hybr_ten} Ben\'{\i}tez Mart\'{\i}nez, F.; Olmedo, J. Primordial tensor modes of the early universe. {\em Phys. Rev. D} {\bf 2016}, {\em 93}, 124008.

\bibitem{dressed3} Agullo, I.; Ashtekar, A.; Nelson, W. The pre-inflationary dynamics of loop quantum cosmology: Confronting quantum gravity with observations. {\em Class. Quant. Grav.} {\bf 2013}, {\em 30}, 085014.

\bibitem{Agullo1} Agullo, I.; Nelson, W.; Ashtekar, A. Preferred instantaneous vacuum for linear scalar fields in cosmological space-times. {\em Phys. Rev. D} {\bf 2015}, {\em 91}, 064051.

\bibitem{NBMmass} Elizaga Navascu\'es, B.; Mart\'{\i}n de Blas, D.; Mena Marug\'an, G.A. Time-dependent mass of cosmological perturbations in the hybrid and dressed metric approaches to loop quantum cosmology. {\em Phys. Rev. D} {\bf 2018}, {\em 97}, 043523.

\bibitem{NBM} Elizaga Navascu\'es, B.; Mart\'{\i}n de Blas, D.; Mena Marug\'an, G.A. The vacuum state of primordial fluctuations in hybrid loop quantum cosmology. {\em Universe} {\bf 2018}, {\em 4}, 98.

\bibitem{Parker} Parker, L. Quantized fields and particle creation in expanding universes. I. {\em Phys. Rev.} {\bf 1969}, {\em 183}, 1057.

\bibitem{Lueders} Lueders, C.; Roberts, J.E. Local quasiequivalence and adiabatic vacuum states. {\em Commun. Math. Phys.} {\bf 1990}, {\em 134}, 29.

\bibitem{Handley} Handley, W.; Lasenby, A.; Hobson, M. Novel quantum initial conditions for inflation. {\em Phys. Rev. D} {\bf 2016}, {\em 94}, 024041.

\bibitem{AG1} Ashtekar, A.; Gupt, B. Quantum gravity in the sky: Interplay between fundamental theory and observations. {\em Class. Quant. Grav.} {\bf 2016}, {\em 34}, 014002.

\bibitem{AG2} Ashtekar, A.; Gupt, B. Initial conditions for cosmological perturbations. {\em Class. Quant. Grav.} {\bf 2017}, {\em 34}, 035004.

\bibitem{deBlas} Mart\'{\i}n de Blas , D.; Olmedo, J. Primordial power spectra for scalar perturbations in loop quantum cosmology. \emph{J. Cosmol. Astropart. Phys. JCAP} {\bf 2016}, {\em 06}, 029.

\bibitem{NMT} Elizaga Navascu\'es, B.; Mena Marug\'an, G.A.; Thiemann, T. Hamiltonian diagonalization in hybrid quantum cosmology. {\em Class. Quant. Grav.} {\bf 2019}, {\em 36}, 185010.

\bibitem{NM} Elizaga Navascu\'es, B.; Mena Marug\'an, G.A. Analytical investigation of pre-inflationary effects in the primordial power spectrum: From general relativity to hybrid loop quantum cosmology. \emph{J. Cosmol. Astropart. Phys. JCAP} {\bf 2021}, {\em 9}, 030.

\bibitem{taveras} Taveras, V. Corrections to the Friedmann equations from loop quantum gravity for a universe with a free scalar field. {\em Phys. Rev. D} {\bf 2008}, {\em 78}, 064072.

\bibitem{PlanckInfla} Akrami, Y. { et al.} [Planck Collaboration]. Planck 2018 results. X. Constraints on inflation. {\emph{Astron. Astrophys.}} {\bf 2020}, {\em 641}, A10.

\bibitem{APS_Ham} Ashtekar, A.; Paw\l{}oski, T.; Sing, P. Quantum nature of the Big Bang: An analytical and numerical investigation. {\em Phys. Rev. D} {\bf 2006}, {\em 73}, 124038.

\bibitem{amb1} Thiemann, T. Quantum spin dynamics (QSD). {\em Class. Quant. Grav,} {\bf 1998}, {\em 15}, 839.

\bibitem{amb2} Assanioussi, M.; Lewandowski, J.; Mäkinen, I. New scalar constraint operator for loop quantum gravity. {\em Phys. Rev. D} {\bf 2015}, {\em
92}, 044042.

\bibitem{amb3} Yang, J.; Ding, Y.; Ma, Y. Alternative quantization of the Hamiltonian in loop quantum cosmology. {\em Phys. Lett. B} {\bf 2009}, {\em 682}, 1.

\bibitem{Engle} Engle, J.; Vilensky, I. Uniqueness of minimal loop quantum cosmology dynamics. {\em Phys. Rev. D} {\bf 2019}, {\em 100}, 121901.

\bibitem{ads} Bentivegna, E.; Paw{\l}owski, T. Anti-de Sitter universe dynamics in loop quantum cosmology. {\em Phys. Rev. D} {\bf 2008},{\em 77}, 124025.

\bibitem{lambd} Paw{\l}owski, T.; Ashtekar, A. Positive cosmological constant in loop quantum cosmology. {\em Phys. Rev. D} {\bf 2012}, {\em 85}, 064001.

\bibitem{Wang1} Li, B.-F.; Singh, P.; Wang, A. Qualitative dynamics and inflationary attractors in loop cosmology. {\em Phys. Rev. D} {\bf 2018}, {\em 98}, 066016.

\bibitem{Wang2} Saeed, J.; Pan, R.; Brown, C.; Clevear, G.; Wang, A. Universal properties of the evolution of the Universe in modified loop quantum cosmology. \emph{arXiv} \textbf{2024},  {\em arXiv:2406.06745}. 

\bibitem{immirzi} Immirzi, G. Real and complex connections for canonical gravity. {\em Class. Quant. Grav.} {\bf 1997}, {\em 14}, L177.

\bibitem{BHLQG1} Meissner, K.A. Black-hole entropy in loop quantum gravity. {\em Class. Quant. Grav.} {\bf 2004}, {\em 21}, 5245.

\bibitem{BHLQG2} Domagala, M.; Lewandowski, J. Black-hole entropy from quantum geometry. {\em Class. Quant. Grav.} {\bf 2004}, {\em 21}, 5233.

\bibitem{Liddle} Liddle, A.R.; Leach, S.M. How long before the end of inflation were observable perturbations produced? {\em Phys. Rev. D} {\bf 2003}, {\em 68}, 103503.

\bibitem{Parker1} Parker, L.; Fulling, S.A. Adiabatic regularization of the energy-momentum tensor of a quantized field in homogeneous spaces. {\em Phys. Rev. D} {\bf 1974}, {\em 9}, 341.

\bibitem{Anderson} Anderson, P.R.; Parker, L. Adiabatic regularization in closed Robertson-Walker universes. {\em Phys. Rev. D} {\bf 1987}, {\em 36}, 2963.

\bibitem{SLE} Mart\'{\i}n-Benito, M.; Neves, R. B.; Olmedo, J. States of low energy in bouncing inflationary scenarios in loop quantum cosmology. {\em Phys. Rev. D} {\bf 2021}, {\em 103}, 123524.

\bibitem{NMY} Elizaga Navascu\'es, B.; Mena Marug\'an, G.A.; Y\'ebana Carrilero, J. Effects of the inflaton potential on the primordial power spectrum in loop quantum cosmology scenarios. {\em Phys. Rev. D} {\bf 2023}, {\em 108}, 083521.

\bibitem{AMV} Alonso-Serrano, A.; Mena Marug\'an, G.A.; Vicente-Becerril, A. Primordial power spectrum in modified cosmology: From thermodynamics of spacetime to loop quantum cosmology. \emph{arXiv}  \textbf{2023}, { arXiv:2307.06813}.

\bibitem{waco} Wu, Q.; Zhu, T.; Wang, A. Nonadiabatic evolution of primordial perturbations and non-Gaussinity in hybrid approach of loop quantum cosmology. {\em Phys. Rev. D} {\bf 2018}, {\em 98}, 103528.

\bibitem{waco2} Zhu, T.; Wang, A.; Cleaver, G.; Kirsten, K.; Sheng, Q. Pre-inflationary universe in loop quantum cosmology. {\em Phys. Rev. D} {\bf 2017}, {\em 96}, 083520.

\bibitem{Abra} Abramowitz, M.; Stegun, I.A. {\em Handbook of Mathematical Functions with Formulas, Graphs, and Mathematical Tables}, revised 9th ed.;  {National Bureau of Standards Applied Mathematics Series No. 55}; U.S. Government Printing Office: Washington, DC, USA, 1972. 

\bibitem{Morris} Agullo, I.; Morris, N.A. Detailed analysis of the predictions of loop quantum cosmology for the primordial power spectra. {\em Phys. Rev. D} {\bf 2015}, {\em 92}, 124040.

\end{thebibliography}
\end{document}